\documentclass[pra,twocolumn,aps,showpacs,tightenlines,superscriptaddress,amsmath,amssymb,showkeys,10pt,nofootinbib]{revtex4-2}
\pdfoutput=1
\usepackage[utf8]{inputenc}
\usepackage{comment}
\usepackage{cancel}
\usepackage{graphicx} 
\usepackage{amsmath,mathtools}
\usepackage{xcolor}
\usepackage{amssymb}
\usepackage{amsfonts, amsthm}
\usepackage{dsfont}

\usepackage{simplewick}
\usepackage[colorlinks=true, citecolor=blue, urlcolor=blue, breaklinks=true]{hyperref} 
\usepackage{booktabs}
\usepackage{adjustbox}
\usepackage{siunitx}

\usepackage[normalem]{ulem}

\usepackage{microtype}

\makeatletter
\def\smalloverbrace#1{\mathop{\vbox{\m@th\ialign{##\crcr\noalign{\kern3\p@}%
  \tiny\downbracefill\crcr\noalign{\kern3\p@\nointerlineskip}
  $\hfil\displaystyle{#1}\hfil$\crcr}}}\limits}
\makeatother

\def\dim{\operatorname{dim}}

\def\>{\rangle} \def\<{\langle}

\def\T+{\mathsf{T}_+}

\mathchardef\mhyphen="2D

\newcommand{\ket}[1]{|#1\rangle}

\date{June 12, 2026}

\begin{document}

\title{GPU-accelerated semidefinite programming for causal games}

\author{Emanuel-Cristian Boghiu}
\email{emanuel.boghiu@bsc.es}
\affiliation{Barcelona Supercomputing Center, Plaça d'Eusebi Güell, 1-3 08034, Barcelona, Spain}
\affiliation{Fakult\"at f\"ur Mathematik, Universit\"at Wien, Oskar-Morgenstern-Platz 1, 1090 Vienna, Austria}

\author{Kyrylo Simonov}
\affiliation{Fakult\"at f\"ur Mathematik, Universit\"at Wien, Oskar-Morgenstern-Platz 1, 1090 Vienna, Austria}

\begin{abstract}
The process matrix formalism describes quantum correlations in scenarios without a fixed causal order between local laboratories. Operational signatures of such correlations can be investigated through causal games. A paradigmatic example is the Guess-Your-Neighbour's-Input game, in which two parties attempt to guess each other's inputs. Correlations compatible with any definite, or probabilistically mixed, causal order cannot achieve a winning probability exceeding $1/2$. The best process-matrix strategy currently known attains a value of approximately $0.6218$ using local dimension $d=5$, while the strongest known dimension-independent upper bound is $0.7592$. In this work, we investigate whether increasing the local dimension beyond $d = 5$ can narrow this gap. To this end, we employ a see-saw optimization scheme in which each step is formulated as a semidefinite program. For scalability, we develop a custom implementation of the SCS solver in which the dominant computational cost, the projection onto the positive-semidefinite cone, is offloaded to a GPU, yielding a six-fold speedup. Using this implementation, we explore local dimensions up to $d = 8$, and we do not find significant improvements over the value at $d=5$. Our results suggest that either qualitatively different strategies are required to approach the known upper bound, or that the bound itself is not tight.

\end{abstract}

\maketitle 

\section{Introduction}\label{sec:intro}

Causality is a key component of our description of the world. Explaining correlations in terms of cause and effect allows us to identify the interventions required to achieve a desired outcome or to predict the consequences of our actions. Determining whether observed correlations admit a given causal explanation is the subject of causal inference~\cite{Pearl_2009}, a field that has developed rigorous tools for reasoning about cause-and-effect relationships from observational data. In this framework, causal relations are represented by directed acyclic graphs, where nodes correspond to random variables, arrows indicate direct causal influences, and acyclicity enforces that causes cannot follow their effects, thus avoiding logical paradoxes such as the grandfather paradox.

Quantum theory challenges this classical notion of causality in several ways. Bell nonlocality~\cite{Brunner_2014} demonstrates that quantum correlations between distant parties cannot be explained by any shared classical common cause, even though the parties cannot signal to one another. This implies that Reichenbach's principle~\cite{reichenbach1991direction}, according to which common causes screen off their effects, must be reformulated in the quantum setting. More radically, quantum theory also allows situations in which causal order itself becomes indefinite. A paradigmatic example is the quantum \texttt{SWITCH}~\cite{PhysRevA.88.022318}, where a control system coherently places two operations in one order or the other. In the context of quantum gravity, where the spacetime metric may itself be a quantum object, one expects this idea to extend to superpositions of geometries and therefore of entire causal structures~\cite{hardy2005probabilitytheoriesdynamiccausal}. These developments raise a fundamental question: can such phenomena be understood within a generalized notion of causality, and can logically consistent physical scenarios exist in the absence of a definite causal order? 

Research on quantum causality can broadly be divided into \textit{bottom-up} and \textit{top-down} approaches. Bottom-up approaches start from operational or interpretational principles and use them to construct notions of quantum causal structure. Examples include quantum causal models~\cite{PhysRevX.7.031021, barrett2020quantumcausalmodels}, which extend the classical notion of a causal network and generalize Reichenbach's common-cause principle to the quantum setting, together with their extensions to cyclic structures~\cite{Barrett_2021}; routed and consistent circuits~\cite{Vanrietvelde2021routedquantum, Vanrietvelde2025consistentcircuits}, which construct logically consistent processes from constrained circuit diagrams; and hierarchies of higher-order operations such as quantum supermaps and quantum-controlled circuits~\cite{Wechs_2021}. These frameworks seek to identify physically meaningful manifestations of indefinite causal order and to understand their operational consequences.

The quantum \texttt{SWITCH} is the most prominent example arising within this bottom-up perspective. It can be viewed as a higher-order transformation that coherently controls the order in which two operations are applied, and occupies the lowest nontrivial level of the hierarchy of quantum-controlled circuits. Treating its causal nonseparability as a resource leads to advantages in communication and computation tasks~\cite{chiribella2009, Chiribella_2012, Chiribella_2013, Araujo2014comp, Araujo2017comp2, PhysRevLett.117.100502, Caleffi2020, Chiribella_2021, Renner2022, Caleffi2023, Liu2024, yin2024magic, zhao2026communicationpowerindefinitecausal, Simonov2026comp} as well as finds applications in metrology \cite{zhao2020quantum, Bavaresco2021, Chapeau2021, Chapeau2022, Liu2023, Chapeau2023, Goldberg2023, Mothe2024} and thermodynamics \cite{felce2020quantum, Guha2020, cao2022quantum, nie2022experimental, simonov2022work, Goldberg2023thermo, dieguez2023thermal, Zhu2023, simonov2025Activation, xue2025anomalousheatflowquantum, wang2025experimentaldemonstrationgenuinequantum, Lisboa2026}. Experimental realizations have been reported~\cite{procopio2015experimental, rubino2017experimental, goswami2018indefinite, wei2019experimental, guo2020experimental}; however, whether these experiments demonstrate genuine indefinite causal order in spacetime remains a matter of ongoing debate~\cite{Paunkovic2020causalorders, Vilasini_2024, Oreshkov_2019}.

While bottom-up approaches focus on physically motivated constructions, top-down approaches seek to characterize the most general correlations compatible with local quantum mechanics and logical consistency, without assuming any underlying spacetime picture or physical realization. A prominent example is the process matrix formalism of Oreshkov, Costa, and Brukner~\cite{oreshkov_quantum_2012}. In this framework, each laboratory performs ordinary quantum operations on incoming and outgoing systems, while no global causal order between laboratories is assumed. Correlations are mediated by a process matrix $W$, which generalizes the role played by quantum states in standard quantum theory. Process matrices that cannot be written as convex mixtures of behaviors compatible with definite causal orders are called \textit{causally nonseparable}, playing a role analogous to entangled states in Bell scenarios.

One of the main strengths of the process matrix formalism is that it allows one to study operational signatures of indefinite causal order. In particular, a strict subclass of process matrices gives rise to correlations that violate \textit{causal inequalities}~\cite{oreshkov_quantum_2012, branciard2016simplest}, that is, whose output statistics lie outside the polytope of correlations compatible with any definite, or probabilistically mixed, causal order between the parties. Violating a causal inequality therefore certifies the absence of a definite causal order in a device-independent way, directly from the observed probabilities and without assuming that they arise from a process matrix (or even from quantum theory), just as violations of Bell inequalities certify nonlocality. This notion of causal indefiniteness is strictly stronger than causal nonseparability: the quantum \texttt{SWITCH}, for example, is causally nonseparable~\cite{araujo_witnessing_2015}, yet its observable correlations always remain inside the causal polytope~\cite{Barrett_2021}. As in the Bell scenario, understanding the maximal possible violation of causal inequalities is therefore a fundamental and largely open problem.

A paradigmatic example is provided by the \textit{Guess-Your-Neighbour's-Input} (GYNI) game~\cite{Almeida_2010, branciard2016simplest}, in which two parties attempt to guess each other's inputs. Any correlation compatible with a definite causal order achieves a winning probability of at most $1/2$, and the same bound holds for arbitrary probabilistic mixtures of causal orders. 
The best process-matrix strategy currently known achieves a winning probability of approximately $0.6218$ using local systems of equal dimension $d = 5$~\cite{branciard2016simplest}, whereas the strongest known dimension-independent upper bound is approximately $0.7592$~\cite{liu2025tsirelson}. Whether this gap reflects a genuine limitation of process matrices or merely the limitations of current numerical methods remains an open question. 

Existing searches for process-matrix strategies typically rely on see-saw heuristics~\cite{branciard2016simplest, Werner_2001}, which require many random restarts and the repeated solution of semidefinite programs (SDPs). In the symmetric bipartite setting, where all local input and output dimensions are equal to $d$, the process matrix is defined on a Hilbert space of dimension $d^4$. As a result, numerical investigations have so far been restricted to relatively small local dimensions, as the size of the SDP variable grows rapidly with $d$. 

The computational challenges encountered in process-matrix optimization reflect a broader trend in semidefinite optimization. Motivated by large-scale applications, e.g., in combinatorial optimization~\cite{boyd1997semidefinite} and optimal power flow~\cite{Lavaei_OPF_2012}, recent years have seen a shift from CPU-based interior-point methods towards first-order algorithms designed for GPU hardware.

One prominent approach is the low-rank Burer--Monteiro factorization~\cite{burer2003nonlinear, burer2005local}, which replaces the positive-semidefinite variable by a nonlinear parametrization $X = FF^\top$ of lower inner dimension. This idea underlies GPU solvers such as cuLoRADS~\cite{han2024acceleratinglowrankfactorizationbasedsemidefinite} and cuHALLaR~\cite{aguirre2025cuhallargpuacceleratedlowrank}, and has proved highly effective for SDPs whose optimal solutions are expected to be low rank. Process matrices, however, need not possess this property, making the applicability of such methods to our setting unclear.

An alternative strategy is to retain the full semidefinite variable while accelerating the computationally dominant operations. In the optimization problems considered here, the principal cost per iteration is the projection onto the positive-semidefinite cone, which requires an eigendecomposition. Among the few GPU implementations of this approach are the interior-point solver CuClarabel~\cite{CuClarabel} and the first-order solver cuADMM~\cite{kang2024fastcertifiabletrajectoryoptimization}. However, both are primarily designed for real symmetric programs, whereas the SDPs from the process matrix formalism are naturally formulated over complex Hermitian matrices. Embedding Hermitian constraints into a real symmetric representation doubles the matrix dimension and can substantially increase the computational cost. Moreover, the current GPU implementation of CuClarabel restricts semidefinite blocks to small blocks of equal size~\cite[Footnote 2]{CuClarabel}, limiting its applicability to the large structured SDPs considered here.

In this work, we develop a custom implementation of the operator-splitting solver SCS~\cite{ocpb_16, odonoghue_21}, tailored to the structure of the process matrix formalism. In the current implementation of SCS (version 3.2.11), GPU acceleration is limited to the affine-projection step, while the complex semidefinite-cone projection is performed on the CPU. We instead offload this projection to the GPU, using single precision during the early stages of the optimization and switching to double precision near convergence, yielding up to a six-fold speedup. As in the standard implementation, we work directly with complex Hermitian variables and therefore avoid the real embedding. Furthermore, the orthonormality of our process-matrix basis reduces the affine-projection step to a single application of the constraint map and its adjoint. We retain standard acceleration techniques, including warm-starting, Anderson acceleration, and over-relaxation, and use the resulting solver to investigate whether increasing the local dimension can substantially improve the best known GYNI winning probability. 

\section{Preliminaries}

Let us consider a bipartite scenario with two local laboratories, Alice's and Bob's. Each laboratory has an incoming and an outgoing quantum system, described by Hilbert spaces $A_{\mathrm{i}},A_{\mathrm{o}}$ for Alice and $B_{\mathrm{i}},B_{\mathrm{o}}$ for Bob. Upon receiving classical inputs $x$ and $y$, the parties implement local quantum operations and produce classical outputs $a$ and $b$, respectively.

For every input $x$, Alice's operation is described by a quantum instrument $\{\Lambda_{a|x}\}_a$, that is, a collection of completely positive maps such that $\sum_a \Lambda_{a|x}$ is trace-preserving. Bob's operation is described, analogously, by an instrument $\{\Lambda_{b|y}\}_b$. Thus, the dynamics inside each laboratory are assumed to obey standard quantum theory.

Throughout this work we employ the Choi representation of the local quantum operations~\cite{de1967linear, jamiolkowski1972linear, choi1975completely}. The Choi operators associated with the instrument elements $\Lambda_{a|x}$ and $\Lambda_{b|y}$ are denoted by 
\[ 
    C_{a|x}\in\mathcal L(A_{\mathrm{i}}\otimes A_{\mathrm{o}}), \qquad C_{b|y}\in\mathcal L(B_{\mathrm{i}}\otimes B_{\mathrm{o}}), 
\] 
respectively, where $\mathcal{L}(\mathcal{H})$ denotes the space of linear operators on the Hilbert space $\mathcal{H}$. Each Choi operator is positive semidefinite, and the condition that the elements of an instrument sum to a quantum channel translates into a set of linear constraints. Their explicit form is given in Appendix~\ref{ap:channel_proj}.

The process matrix formalism provides an operational framework for describing correlations between the local laboratories without assuming an underlying definite causal order. One postulates a probability assignment 
\begin{equation} 
    P(C_{a|x}, C_{b|y}) = p(ab|xy), 
\end{equation} 
which associates probabilities with pairs of local instrument elements. Requiring this assignment to be compatible with classical randomization and coarse-graining of local operations implies that $P$ is bilinear in its arguments. A detailed derivation is presented in Appendix~\ref{ap:from_oper_to_born}.

By the universal property of the tensor product, such a bilinear map induces a linear functional on the tensor product of the local operator spaces. Since these spaces are finite dimensional, this functional can be represented through the Hilbert--Schmidt inner product by a Hermitian operator $W \in \mathcal{L}(A_{\mathrm{i}}\otimes A_{\mathrm{o}}\otimes B_{\mathrm{i}}\otimes B_{\mathrm{o}})$. The probability rule therefore takes the generalized Born form
\begin{equation}
    p(ab|xy)
    =
    \operatorname{tr}\!\left[
        W\,(C_{a|x}\otimes C_{b|y})
    \right].
    \label{eq:born_rule}
\end{equation}

\begin{figure}
    \centering
    \includegraphics[width=0.7\linewidth]{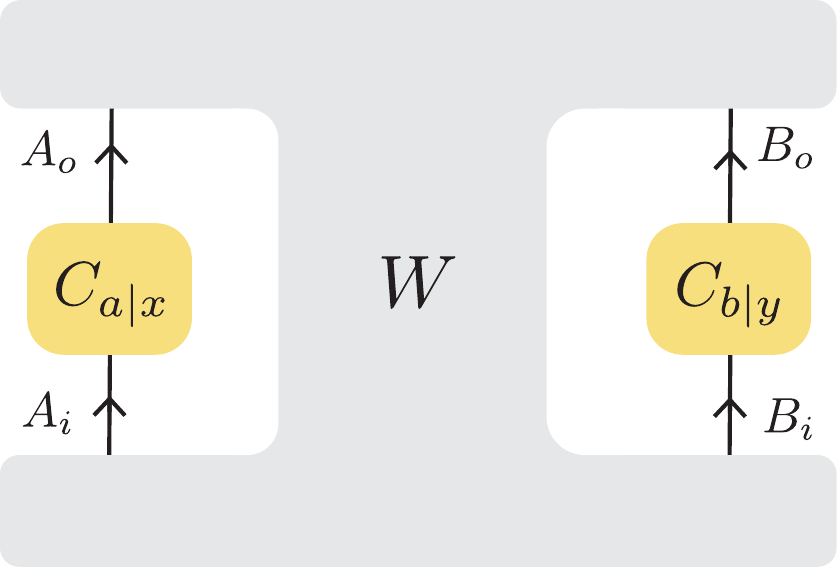}
    \caption{Schematic representation of the bipartite scenario within the process matrix formalism. Given local instruments $C_{a|x}$ and $C_{b|y}$, the process matrix $W$ assigns joint probabilities $p(ab|xy)=\operatorname{tr}[W(C_{a|x}\otimes C_{b|y})]$ for all settings $x,y$ and outcomes $a,b$.}
    \label{fig:bipartite_scenario}
\end{figure}

A Hermitian operator $W$ is a \textit{process matrix} if Eq.~\eqref{eq:born_rule} defines valid probabilities for all choices of local quantum instruments, that is,
\begin{equation} 
    p(ab|xy) \geq 0, \qquad \sum_{a,b} p(ab|xy) = 1,
\end{equation}
for all values of $a,b,x,y$. The normalization condition imposes linear constraints on $W$, whose explicit form is given in Appendix~\ref{ap:proc_projector}. If arbitrary shared ancillary entanglement between the laboratories is allowed, positivity of the induced functional is equivalent to the semidefinite constraint 
\[
    W \succeq 0,
\]
as shown in Refs.~\cite{oreshkov_quantum_2012,araujo_witnessing_2015}. Fig.~\ref{fig:bipartite_scenario} illustrates the bipartite process-matrix scenario.

The generalized Born rule associates with each process matrix and each choice of local instruments a conditional probability distribution $p(ab|xy)$, which we call a behaviour. A behaviour is causal if it can be written as a convex mixture of behaviours compatible with the two definite causal orders,
\begin{equation}\label{eq:causal_separability}
    p(ab|xy) = q\,p^{A\prec B}(ab|xy) + (1-q)\,p^{B\prec A}(ab|xy). 
\end{equation}
for some $q\in[0,1]$~\cite{araujo_witnessing_2015, Oreshkov_2016, branciard2016simplest}. Here $p^{A\prec B}$ denotes a behaviour compatible with Alice preceding Bob, whereas $p^{B\prec A}$ denotes a behaviour compatible with Bob preceding Alice.

At the level of observed correlations, compatibility with $A\prec B$ requires Alice's marginal distribution to be independent of Bob's input, while compatibility with $B\prec A$ requires Bob's marginal distribution to be independent of Alice's input:
\begin{align}
     \textstyle\sum_b p^{A\prec B}(ab|xy) &= p^{A\prec B}(a|x), \\
     \textstyle\sum_a p^{B\prec A}(ab|xy) &= p^{B\prec A}(b|y).
\end{align}
Behaviours that do not admit the decomposition~\eqref{eq:causal_separability} are called noncausal. This notion is distinct from causal nonseparability of process matrices: a causally nonseparable process need not generate a behaviour that violates a causal inequality.

Causal inequalities are linear constraints satisfied by all causal behaviours. Equivalently, they can be formulated as causal games. In such a game, Alice and Bob receive inputs $x$ and $y$, produce outputs $a$ and $b$, and are scored according to a payoff function $\omega_{abxy}$. The average payoff of a behaviour is
\[
    \omega(p)
    =
    \sum_{a,b,x,y}
    \omega_{abxy}\,p(ab|xy).
\]
If this value exceeds the maximum over all causal behaviours, then the behaviour violates a causal inequality and is therefore noncausal.

The main example considered in this work is the GYNI game~\cite{Almeida_2010, branciard2016simplest}. Alice and Bob receive uniformly distributed binary inputs $x,y \in \{0,1\}$ and produce binary outputs $a,b \in \{0,1\}$. They win whenever each party correctly guesses the other party's input, i.e., when $a = y$ and $b = x$. The corresponding winning probability is:
\begin{equation}
    \omega^{\mathrm{GYNI}}(p)
    =
    \frac{1}{4}
    \sum_{x,y,a,b}
    \delta_{a,y}\,\delta_{b,x}\,p(ab|xy).
\end{equation}

For every causal behaviour, $\omega^{\mathrm{GYNI}}\leq 1/2$. Indeed, if Alice precedes Bob, then Bob may learn Alice's input and guess it perfectly, but Alice cannot learn Bob's input and can do no better than random guessing. The opposite situation holds for the order $B\prec A$, and convex mixtures of the two orders cannot improve this bound.

Process matrices can generate behaviours that exceed this causal bound. The best known process-matrix strategy achieves $\omega^{\mathrm{GYNI}} \approx 0.6218$ using local dimension $\dim A_{\mathrm{i}} = \dim A_{\mathrm{o}} = \dim B_{\mathrm{i}} = \dim B_{\mathrm{o}} = 5$. Whether this value is optimal remains unknown. Perfect success is impossible~\cite{bavaresco2024indefinite}, while the strongest currently known Tsirelson-type (i.e., dimension-independent) upper bound is 
\begin{equation}\label{eq:tsirelson}
    \omega^{\mathrm{GYNI}}\leq 0.7592,
\end{equation}
as established in Ref.~\cite{liu2025tsirelson}.

\section{Methods}

Our objective is to investigate whether the Tsirelson-type upper bound~\eqref{eq:tsirelson} can be approached, or even saturated, by increasing the local dimension beyond the value $d = 5$ considered in Ref.~\cite{branciard2016simplest}. Throughout this work we restrict attention to the symmetric bipartite setting,
\[
    \dim A_{\mathrm{i}} = \dim A_{\mathrm{o}} = \dim B_{\mathrm{i}} = \dim B_{\mathrm{o}} = d.
\]

The optimization is performed by means of a see-saw algorithm, following the approach of Ref.~\cite{branciard2016simplest}. The procedure alternates between optimizing the local instruments for a fixed process matrix and optimizing the process matrix for fixed local instruments. Each individual optimization is a SDP, whereas the joint optimization problem is non-convex. Consequently, convergence to a global optimum is not guaranteed. A detailed description of the see-saw algorithm is given in Appendix~\ref{app:see-saw}.

The optimization over local instruments is carried out using a sparse affine parametrization in an orthonormal Hermitian basis adapted to the trace-preserving constraint. Within this representation, every instrument element decomposes naturally into three sectors: a baseline sector, proportional to the identity on the input and output spaces; an effect-like sector, encoding the corresponding element of the positive operator-valued measure (POVM) induced by the instrument; and a transformation-like sector, describing the nontrivial input-output dependence of the operation. This parametrization is tight, in the sense that it introduces no redundant variables, while incorporating the normalization constraints exactly. The construction and the corresponding SDP formulation are presented in Appendix~\ref{ap:affine_instrument}.

For the optimization over process matrices, we use an orthonormal traceless basis $\{F_\mu\}_\mu$ adapted to the linear constraints defining the process-matrix subspace, following the characterization of Ref.~\cite{oreshkov_quantum_2012}. Choosing the generalized Gell--Mann basis~\cite{bertlmann2008bloch} for each local subsystem yields a sparse parametrization in which both the process constraints and the trace normalization are incorporated directly into the SDP variables.

Choosing the generalized Gell--Mann basis for each local subsystem naturally partitions the local operator basis elements into baseline, effect-like, and transformation-like sectors, mirroring the corresponding decomposition of the instrument Choi operators. The process constraints determine which tensor-product combinations of these local sectors are allowed to appear in the expansion of a process matrix. In particular, every transformation-like factor must be accompanied by at least one effect-like factor belonging to another laboratory. Tensor-product basis elements containing transformation-like factors coupled only to baseline sectors, or exclusively to other transformation-like factors, survive the complete outcome summation and therefore lead to instrument-dependent normalizations. Such terms are incompatible with the process constraints and are excluded from the process-matrix subspace.

The sparse construction extends naturally to the $N$-partite setting. In particular, the sector decomposition yields a closed expression for the dimension of the admissible process subspace. For equal local dimension $d$,
\begin{equation}
    \frac{|\mathcal A_N|}{d^{4N}}
    =
    1-
    \left(
        1-\frac{1}{d^2}+\frac{1}{d^4}
    \right)^N
    +
    \frac{1}{d^{4N}} .
\end{equation}
For fixed $N$, the relative size of the process subspace vanishes as $d \rightarrow \infty$, whereas for fixed $d$ it approaches the full operator space as $N \rightarrow \infty$. Consequently, for scenarios involving few parties and large local dimensions, an allowed-basis parametrization is the most economical implementation. By contrast, for many low-dimensional parties, it may be advantageous to impose the forbidden components as explicit linear constraints, leading to SDPs with fewer optimization variables. Further details are provided in Appendix~\ref{ap:sparse_process_basis}.

In practice, the dominant computational cost of the see-saw algorithm arises from the optimization over process matrices, which involves large semidefinite variables together with numerous linear constraints. We investigated alternative strategies, including a joint gradient-based optimization over all variables, but found the combination of the see-saw heuristic with a customized implementation of the SCS solver~\cite{ocpb_16, odonoghue_21} to provide the most reliable performance. A brief discussion of the gradient-based approach is included in Appendix~\ref{sec:other_methods}.

\begin{table*}[t]
    \centering
    \renewcommand{\arraystretch}{1.15}
    \setlength{\tabcolsep}{6pt}
    \begin{adjustbox}{max width=1.5\columnwidth}
    \begin{tabular}{cccccccc}
        \toprule
        $d$
        & $2$ & $3$ & $4$ & $5$ & $6$ & $7$ & $8$  \\
        \midrule
        $\omega^{\mathrm{GYNI}}$
        & $0.5694$ & $0.6104$ & $0.6217$ & $0.6218$ & $0.6219$
        & $0.6219$ & $0.6218(9)$  \\
        Ref.~\cite{branciard2016simplest}
        & $0.5694$ & $0.6104$ & $0.6201$ & $0.6218$
        & -- & -- & --  \\
        \bottomrule
    \end{tabular}
    \end{adjustbox}
    \caption{Maximum winning probability $\omega^{\mathrm{GYNI}}$ obtained by the see-saw optimization over process-matrix correlations as a function of the local dimension $d$. The second row reports the best values previously obtained in Ref.~\cite{branciard2016simplest}.}
    \label{tab:seesaw-results}
\end{table*}

SCS is a first-order operator-splitting algorithm in which each iteration consists of two principal steps: a projection onto the positive-semidefinite cone, obtained through an eigendecomposition, and a projection onto the affine constraint set, which reduces to the solution of a linear system. The solver provides both a direct implementation based on sparse matrix factorization and an indirect implementation based on the conjugate-gradient method applied to the reduced system
\begin{align}
    (I + A^\top A) z_x &= r_x - A^\top r_y, \\
    z_y &= a_y + A z_x,
\end{align}
where $A$ is obtained by stacking the real vectorizations of the basis elements $\{F_\mu\}_\mu$. 

In the present setting, the orthonormality of the generalized Gell--Mann basis implies $A^\top A = I$, so that the affine projection admits a closed-form expression involving a single application of $A$ and $A^\top$. Consequently, the distinction between the direct and indirect implementations becomes immaterial.

The dominant cost per iteration is therefore the eigendecomposition required for the semidefinite-cone projection, which, in the implementations of SCS up to version 3.2.11, is performed on the CPU. We implement a customized version in Python that offloads this computation to the GPU through JAX~\cite{jax2018github}. In addition, we employ a mixed-precision strategy, performing the initial iterations in single precision and switching to double precision only near convergence. Empirically, this approach provides up to a six-fold speedup against the native C implementation of SCS while maintaining numerical accuracy. An overview of the SCS algorithm and a detailed description of our implementation together with benchmarks are given in Appendix~\ref{app:scs}.

\section{Results}

The results of the see-saw optimization are summarized in Table~\ref{tab:seesaw-results}. The numerical study was performed in two stages.

In an exploration stage, the see-saw algorithm was run for many random initializations at a looser tolerance to survey the optimization landscape. The best solution was then used to warm-start a refinement stage at a stricter tolerance. Each iteration of the see-saw algorithm involves SDP optimizations over the instruments and the process matrix, and the see-saw optimization terminates once the average payoff changes by less than $\epsilon$ between successive steps. Throughout our study, the process matrix SDP and the see-saw loop use the same tolerance, while the instrument SDP was solved at a precision of $\epsilon=10^{-8}$. For $d \leq 5$, the optimization was performed on the CPU, using $\epsilon=10^{-6}$ for exploration and $10^{-7}$ for refinement, while for $d \geq 6$, we used an NVIDIA Hopper GPU with 64\,GB of memory, with $\epsilon=10^{-5}$ for exploration and $\epsilon=10^{-6}$ for refinement. The looser tolerance at large local dimension was chosen to favor exploration of the optimization landscape. All computations were performed on the MareNostrum5 supercomputer~\cite{banchelli2025introducing}, and for each value of the local dimension, we ran the see-saw algorithm for the equivalent of one month of wall-clock time on a single machine. 

After refinement, the resulting process matrix and local instruments were subsequently corrected so as to satisfy the positive-semidefinite condition to machine precision by convexly mixing them with, respectively, the trivial process and white-noise instruments. The details of this post-processing procedure are given in Appendices~\ref{ap:ch_postprocessing} and \ref{ap:proc_postprocessing}. Owing to the parametrizations introduced in the previous section, all remaining affine constraints are satisfied automatically up to numerical precision.

The winning probability is observed to increase with the local dimension, although the improvement becomes marginal beyond $d = 5$. For $d=6$ and $d=7$, we obtain a winning probability of $0.6219$ while at $d=8$, we found a marginally smaller value of $0.62189$. Naturally, a process matrix of dimension $d$ can reproduce any strategy available at a smaller dimension, therefore the optimal winning probability cannot decrease with dimension. The fact that our $d=8$ value does not exceed those found at $d=6$ and $d=7$ reflects a less exhaustive search of the optimization landscape, indicating that we may require more than one month of wall-clock time to recover, or improve upon, the strategies found at lower local dimension. The same effect is already apparent at $d=4$, where we obtain a strategy whose winning probability exceeds the value reported in Ref.~\cite{branciard2016simplest}, namely $0.6217 > 0.6201$. Taken together, these observations highlight the sensitivity of the see-saw heuristic to the number of random initialization and suggests that even relatively low-dimensional instances may not yet have been explored exhaustively.

Despite the improved lower bound, the growth of the winning probability for $d \geq 5$ is remarkably slow. Our numerical evidence therefore suggests that increasing the local dimension alone may not be sufficient to approach the Tsirelson-type upper bound \eqref{eq:tsirelson}. Whether this gap reflects an intrinsic limitation of process-matrix correlations or the limitations of current optimization techniques remains an open question.

\section{Conclusions and Outlook}

We have investigated whether increasing the local dimension can significantly improve the maximal known process-matrix violation of the GYNI causal inequality. To this end, we developed a GPU-accelerated implementation of the SCS semidefinite-programming solver tailored to the structure of the process matrix formalism, combining sparse affine parametrizations with a customized treatment of the positive-semidefinite projection. Using this framework, we explored symmetric bipartite process matrices in local dimensions up to $d = 8$ obtaining an extremely slow growth of the winning probability beyond $d = 5$. Our numerical evidence therefore suggests that increasing the local dimension alone may not be sufficient to approach the currently best known Tsirelson-type upper bound.

The results obtained in this work suggest several directions for further investigation, both from the perspective of indefinite causal order and of large-scale semidefinite optimization.

\paragraph{Improved upper bounds.} The Tsirelson-type upper bound of Ref.~\cite{liu2025tsirelson} for the GYNI inequality may not be tight. Developing stronger relaxations for the set of process-matrix correlations is therefore an important open problem. Ideally, such relaxations would form a convergent hierarchy analogous to the Navascués--Pironio--Acín (NPA) hierarchy for Bell nonlocality~\cite{Navascu_s_2008}. A promising candidate is the recently introduced semidefinite block-moment matrix relaxation~\cite{dalessandro2026semidefiniteblockmatrixrelaxationscomputing}, which extends NPA-type constructions to a broader class of device-dependent constraints, although convergence guarantees are presently unknown. This framework has already found applications in entanglement detection and related problems. Whether it, or an appropriate adaptation thereof, can be used to obtain tighter bounds on the violation of causal inequalities is, in our view, a worthwhile direction for future work.

\paragraph{Approximate semidefinite-cone projections.} A useful property of operator-splitting methods is their robustness to inexact subproblem solutions: convergence is preserved provided the accumulated errors remain summable~\cite{ocpb_16, Eckstein_1992}. This opens the possibility of replacing the exact projection onto the positive-semidefinite cone, which dominates the computational cost of each iteration, by a cheaper approximation. A recent proposal avoids the eigendecomposition entirely~\cite{kang2025factorization}. Since projection onto the positive-semidefinite cone amounts to replacing negative eigenvalues by zero, the corresponding spectral map can be approximated by a composite matrix polynomial constructed solely from matrix--matrix products, with coefficients optimized to reproduce the eigenvalue-thresholding operation. Because this construction acts only on the spectrum of the matrix, it extends directly to the complex Hermitian operators arising in the present work. Such factorization-free projections are particularly well suited to GPU architectures and low-precision arithmetic, and may enable the optimization of GYNI strategies, and semidefinite programs more broadly, in substantially larger dimensions.

\paragraph{Exploiting symmetry.} The GYNI game is invariant under a finite group of relabelings of the parties, inputs, and outputs. Although a symmetric optimization problem does not have necessarily to admit a symmetric optimum, exploiting these symmetries may nevertheless provide significant computational advantages. For example, symmetry reduction may decrease the number of optimization variables or decompose the positive-semidefinite constraint into smaller blocks. Such reductions could make it possible to explore larger local dimensions and help determine whether the slow growth of the GYNI winning probability observed in this work persists.

\paragraph{Multipartite scenarios and higher-order operations.} Although the present work focuses on a single bipartite causal inequality, the methods developed here extend naturally to multipartite settings, where the causal polytope possesses a much richer facet structure~\cite{Abbott_2016} and the dimension of the process matrix grows exponentially with the number of parties. More generally, the same ideas apply to higher-order quantum operations that are characterized by affine constraints generated by trace-and-replace maps, yielding orthogonal projections onto the feasible affine subspace~\cite{Bisio2019, Milz_2024, taranto2025higherorderquantumoperations, apadula2026higherordertransformationsbidirectionalquantum} and therefore requiring no auxiliary linear solve within SCS. The dimension of the corresponding optimization problems nevertheless grows exponentially with the number of slots, leaving the positive-semidefinite projection as the dominant computational cost. Such problems are therefore natural candidates for the GPU-accelerated operator-splitting techniques developed in this work.

\section{Code availability}
The SCS developed, see-saw code and scripts to validate the strategies can be found at:\\

\indent\indent\hspace{-0.3em} \url{https://github.com/ecboghiu/ico-jax-scs} \\

\section{Acknowledgements}
E.C.B. and K.S. thank Jessica Bavaresco, Andreas Leitherer, Simon Milz, and Marco T\'ulio Quintino for fruitful discussions. Claude Opus 4.6 was used to assist with parts of the code development and text editing. This research was funded in whole or in part by the Austrian Science Fund (FWF) 10.55776/PAT4559623. For open access purposes, the author applied a CC BY public copyright license to any author-accepted manuscript version arising from this submission.

\bibliographystyle{apsrev4-1}
\bibliography{biblio}

\appendix

\section{Quantum instrument constraints and parametrizations}
\label{ap:choi_matrix_sdpconstrains}

In this appendix, we describe the characterization of quantum instruments and the corresponding SDP formulation used throughout this work.

We denote Alice's input and output Hilbert spaces by $A_\mathrm{i},A_\mathrm{o}$, and Bob's by $B_{\mathrm{i}},B_{\mathrm{o}}$. We write $d_X=\dim X$, and employ the Choi--Jamio\l kowski isomorphism~\cite{de1967linear,jamiolkowski1972linear, choi1975completely}, under which a linear map 
\[ 
    \Lambda: \mathcal{L}(A_{\mathrm{i}}) \rightarrow \mathcal{L}(A_{\mathrm{o}})
\]
is represented by the Choi operator 
\[ 
    C \coloneqq \sum_{i,j} |i\rangle\langle j| \otimes \Lambda[|i\rangle\langle j|], 
\]
acting on $A_{\mathrm{i}}\otimes A_{\mathrm{o}}$. Here $\mathcal{L}(\mathcal{H})$ denotes the space of linear operators on the Hilbert space $\mathcal{H}$. The action of the map on a quantum state is recovered through
\[ 
    \Lambda(\rho) = \operatorname{tr}_{A_{\mathrm{i}}} \left[ C (\rho^\top\otimes\mathds{1}_{A_{\mathrm{o}}}) \right], 
\]
where $(\cdot)^\top$ denotes transposition in the computational basis.

For any subsystem $X$, we use the trace-and-replace notation,
\begin{equation} \label{eq:trace_and_replace}
    {}_X M
    \coloneqq
    \frac{\mathds{1}_X}{d_X}\otimes \operatorname{tr}_X M ,
\end{equation}
with the tensor factors understood in the appropriate order. The notation extends directly to several subsystems, e.g., ${}_{A_{\mathrm{o}}B_{\mathrm{o}}}M$.

\subsection{Standard SDP formulation of quantum instruments}  \label{ap:channel_proj}

For each input $x=0,\ldots,n_x{-}1$, Alice's instrument consists of a collection of completely positive maps $\{\Lambda_{a|x}\}_{a=0}^{n_a-1}$ such that $\sum_a\Lambda_{a|x}$ is trace-preserving, where $n_a$ denotes the number of classical outcomes. In the Choi representation, these conditions become
\begin{align*}
    C _{a|x}  &\succeq 0,
    \qquad \forall a,x,
    \\
    \operatorname{tr}_{A_{\mathrm{o}}}
        [\textstyle\sum_a C _{a|x}] 
    &=
    \mathds{1}_{A_{\mathrm{i}}},
    \qquad \forall x .
    \label{eq:qinstrument_constraints}
\end{align*}

Defining $S_x \coloneqq \sum_a C_{a|x}$, the trace-preserving condition can equivalently be written as~\cite{Milz2024characterising}:
\begin{align*}
    P_{\mathrm{ch}}[S_x] &= S_x ,
     \\
    \operatorname{tr}[S_x]  &= d_{A_{\mathrm{i}}},
\end{align*}
where
\begin{equation*}
    P_{\mathrm{ch}}[M]
    \coloneqq
    M-{}_{A_{\mathrm{o}}}M+{}_{A_{\mathrm{i}}A_{\mathrm{o}}} M . 
\end{equation*}
Indeed, $P_{\mathrm{ch}}[S_x] = S_x$ is equivalent to ${}_{A_{\mathrm{o}}}S_x = {}_{A_{\mathrm{i}}A_{\mathrm{o}}} S_x$, which, together with $\operatorname{tr}[S_x] = d_{A_{\mathrm{i}}}$ is equivalent to $\operatorname{tr}_{A_{\mathrm{o}}}[S_x] = \mathds{1}_{A_{\mathrm{i}}}$. Bob's constraints are identical, with $A_{\mathrm{i}}, A_{\mathrm{o}}, a, x$ replaced by $B_{\mathrm{i}}, B_{\mathrm{o}}, b, y$.

The optimization over Alice's instruments can therefore be written as the SDP
\begin{align}
    \underset{\{C_{a|x}\}}{\mathrm{max}}
    \quad&
    \sum_{a,x}
    \operatorname{tr}\!\left[
        B_{a|x} C _{a|x} 
    \right] \label{eq:sdp_instrument}
    \\ \nonumber
    \mathrm{s.t.}
    \quad&
    C_{a|x} \succeq0,
    \quad \forall a,x,
    \\ \nonumber
    & S_x = \textstyle\sum_a C_{a|x}, \quad \forall x,
    \\  \nonumber
    &
    P_{\mathrm{ch}}[S_x]
    =
    S_x,
    \quad \forall x,
    \\ \nonumber
    &
    \operatorname{tr}
        [S_x]
    =
    d_{A_{\mathrm{i}}},
    \quad \forall x.
\end{align}
Here, $\{B_{a|x}\}_{a,x}$ are fixed Hermitian operators specifying the objective function. An analogous formulation applies to Bob's instruments.

\subsection{Sparse affine parametrization of quantum instruments} \label{ap:affine_instrument}

In practice, SDP solvers impose constraints only up to numerical precision. For instance, with MOSEK~\cite{aps2022mosek, Andersen_2000} the feasibility tolerance is typically of order $10^{-8}$, whereas machine precision is of the order $10^{-16}$. Although such accuracy is generally sufficient for optimization, the resulting matrices need not correspond exactly to valid quantum instruments. To obtain objects that satisfy the affine constraints exactly and the positivity constraint to machine precision, we reformulate SDP~\eqref{eq:sdp_instrument} using a sparse affine parametrization adapted to the range of the channel projector. In this representation, the trace-preserving constraints are enforced by construction and therefore require no explicit SDP constraints. If the resulting operators are not positive semidefinite to machine precision, positivity is restored by a convex mixture with a depolarizing instrument, which preserves the trace-preserving property.

To construct a basis adapted to the instrument constraints, we represent an instrument $\{\Lambda_{a|x}\}_a$, for each fixed input $x$, as a single quantum-classical channel obtained by adjoining a classical outcome register $R$,
\begin{equation}
    \Lambda_{x}[\rho] = \sum_a \Lambda_{a|x}[\rho] \otimes |a\rangle\!\langle a|_R,
\end{equation}
where the register dimension equals the number of instrument outcomes, $d_R = n_a$. The corresponding Choi operator is given by
\begin{align}
    \nonumber C_x &= \sum_{ij} |i\rangle\langle j|_{A_{\mathrm{i}}} \otimes \sum_a \Lambda_{a|x}[|i\rangle\langle j |_{A_{\mathrm{i}}}] \otimes |a\rangle\!\langle a|_R \\
    &= \sum_a C_{a|x} \otimes |a\rangle\!\langle a|_R.
\end{align}
The Choi operator of the $a$-th instrument element is recovered by projecting onto the classical register state $|a\rangle\!\langle a|_R$ and tracing out the register:
\begin{equation}
    C_{a|x} = \operatorname{tr}_R [(\mathds{1} \otimes |a\rangle\!\langle a|_R)\, C_x].
\end{equation} 
For a quantum instrument, $C_x$ is block diagonal in the classical register, as shown above. However, to construct an affine basis for instrument elements, it is convenient to first characterize the larger affine space of Choi operators $C_x\in\mathcal L(A_{\mathrm{i}}\otimes A_{\mathrm{o}}\otimes R)$ corresponding to arbitrary completely positive trace-preserving (CPTP) maps from $A_{\mathrm{i}}$ to $A_{\mathrm{o}}' = A_{\mathrm{o}}\otimes R$.

We follow the characterization of CPTP maps given in Ref.~\cite[Suppl.]{oreshkov_quantum_2012} and, for notational simplicity, suppress the input label $x$ throughout this discussion. Since the projector $P_{\mathrm{ch}}$ is built from trace-and-replace maps acting on local subsystems, it is convenient to employ local operator bases that separate the identity component from the traceless directions. For each system $X$, let $\{\sigma^X_\alpha\}_\alpha$ be an orthonormal Hermitian basis of $\mathcal{L}(X)$ satisfying:
\begin{align} \label{eq:orthonormal_traceless_basis}
    \sigma^X_0 &=\frac{\mathds{1}_X}{\sqrt{d_X}}, \\
    \operatorname{tr}[\sigma^X_i] &= 0\quad(i\geq1), \\
    \operatorname{tr}[\sigma^X_\alpha\sigma^X_\beta] &= \delta_{\alpha\beta}.
\end{align}
Greek indices $\alpha,\beta,\gamma,\ldots$ run over the entire basis, including the identity element (corresponding to the index value $0$), whereas Latin indices $i,j,k,\ldots$ are reserved for the traceless basis elements. In this representation, a trace-and-replace map simply removes all traceless components on the subsystem being replaced and retains only the identity component. Consequently, the channel constraint becomes a condition on the expansion coefficients. 
We may expand
the Choi operator on input space $A_{\mathrm{i}}$ and output space $A'_{\mathrm{o}} = A_{\mathrm{o}}\otimes R$ as:
\begin{equation}
    C
    =
    \sum_{\alpha,\beta}
    r_{\alpha,\beta}\,
    \sigma^{A_{\mathrm{i}}}_{\alpha}
    \otimes
    \sigma^{A'_{\mathrm{o}}}_{\beta}.
    \label{eq:instrument_basis_expansion}
\end{equation}
In the numerical implementation, we use the identity together with the generalized Gell--Mann basis~\cite{bertlmann2008bloch}, which provides a sparse orthonormal basis of Hermitian operators:
\begin{widetext}
\begin{multline} \label{eq:gellmann_basis}
     \{ \sigma_\alpha^X\}_\alpha = \underbrace{\left\{ \frac{\mathds{1}}{\sqrt{d_X}} \right\}}_{\sigma_0^X} \cup \Biggl\{\sqrt{\frac{1}{(l+1)(l+2)}} \Biggl(\sum_{j=0}^{l} |j\rangle\!\langle j| 
      - (l+1) |l{+}1\rangle\!\langle l{+}1|\Biggr)\Biggr\}_{0\leq l \leq d_X-2}  
     \cup \\  \cup \left\{ \frac{|j\rangle\!\langle k| + |k\rangle\!\langle j|}{\sqrt{2}} \right\}_{0\leq j<k \leq d_X{-}1} 
     \cup \left\{ -i\frac{|j\rangle\!\langle k| - |k\rangle\!\langle j|}{\sqrt{2}} \right\}_{0\leq j<k \leq d_X{-}1}.
\end{multline}
\end{widetext}

The channel constraint $P_{\mathrm{ch}}[C] = C$ is equivalent to ${}_{A'_{\mathrm{o}}}C = {}_{A_{\mathrm{i}}A'_o}C$. Since a trace-and-replace map retains only the identity component on the subsystem being replaced, we obtain:
\begin{align*}
     {}_{A'_{\mathrm{o}}} C&= \sum_\alpha r_{\alpha,0}\,\, \sigma_\alpha^{A_{\mathrm{i}}} \otimes \sigma_0^{A'_{\mathrm{o}}},  \\ 
     {}_{A_{\mathrm{i}}A'_o} C &= \frac{\operatorname{tr}(C)}{\sqrt{d_{A_{\mathrm{i}}} d_{A'_{\mathrm{o}}}}} \,\,\sigma_0^{A_{\mathrm{i}}} \otimes \sigma_0^{A'_{\mathrm{o}}}, 
\end{align*}
where
\[
    \operatorname{tr}\left[\sigma^{A'_{\mathrm{o}}}_{\beta}\right] = \sqrt{d_{A'_{\mathrm{o}}}}\,\delta_{\beta 0}, \qquad     \frac{\mathds{1}_{A'_{\mathrm{o}}}}{{d_{A'_{\mathrm{o}}}}} = \frac{\sigma_0^{A'_{\mathrm{o}}}}{\sqrt{d_{A'_{\mathrm{o}}}}}.
\]
Hence the channel constraint imposes:
\begin{equation} \label{eq:channel_constraint_coefficients}
   r_{0,0} = \frac{\operatorname{tr}[C]}{\sqrt{d_{A_{\mathrm{i}}}d_{A'_{\mathrm{o}}}}}, \qquad r_{i,0} = 0 \quad \forall i\geq1 .
\end{equation}
The channel constraint $\operatorname{tr}[C] = d_{A_{\mathrm{i}}}$ therefore fixes the identity component, while the coefficients $r_{i,0}$ are forced to vanish. These are precisely the coefficients that survive the partial trace over the output system and would generate deviations from the identity operator required for trace normalization.
All remaining coefficients vanish under the partial trace over the output and therefore remain unconstrained.

It follows that the Choi operator of a CPTP map admits the affine parametrization (we now reintroduce the input label $x$):
\begin{equation} \label{eq:cptp_param}
    C_x = F_0 + \sum_{\alpha\geq0,j\geq1} r^x_{\alpha,j} F_{\alpha,j},
\end{equation}
where $r^x_{\alpha,j} \in \mathbb{R}$ for all $x$, and
\begin{align*}
    F_0
    &=
    \sqrt{\frac{d_{A_{\mathrm{i}}}}{d_{A'_{\mathrm{o}}}}}\,
    \sigma^{A_{\mathrm{i}}}_0\otimes\sigma^{A'_{\mathrm{o}}}_0, \\
    \{F_{\alpha,j}\}_{\alpha\geq0,j\geq1}
    &=
    \left\{
        \sigma^{A_{\mathrm{i}}}_\alpha\otimes\sigma^{A'_{\mathrm{o}}}_j
    \right\}_{\alpha\geq0,\;j\geq1}.
\end{align*}
Since every $F_{\alpha,j}$ is traceless, the trace normalization is entirely contained in the fixed component $F_0$. 

To recover the individual instrument elements, we decompose the basis elements of
$A'_{\mathrm{o}} = A_{\mathrm{o}}\otimes R$ appearing in~\eqref{eq:cptp_param} into tensor products of local basis elements on $A_{\mathrm{o}}$ and the outcome register $R$. Since the identity component has already been isolated in the fixed term $F_0$, the remaining basis elements satisfy:
\begin{align*}
    \left\{ \sigma_j^{A'_{\mathrm{o}}} \right\}_{j\geq1}
    &=
    \left\{
        \sigma_\beta^{A_{\mathrm{o}}} \otimes \sigma_\gamma^R
    \right\}_{(\beta,\gamma)\neq(0,0)}.
\end{align*}
That is, each index $j$ labeling a traceless basis element of $A'_{\mathrm{o}}$ is identified with a pair of local indices $(\beta,\gamma)$, where $\beta$ labels the basis elements of $A_{\mathrm{o}}$ and $\gamma$ labels those of the register $R$.

Projecting the register onto the computational basis state $|a\rangle\!\langle a|$ and tracing out $R$ on both sides of~\eqref{eq:cptp_param} yields the affine basis for the instrument element $C_{a|x}$:
\begin{align}
    \nonumber F_0^a &= \frac{1}{n_a} \sqrt{\frac{d_{A_{\mathrm{i}}}}{d_{A_{\mathrm{o}}}}} \,\sigma_0^{A_{\mathrm{i}}} \otimes \sigma_0^{A_{\mathrm{o}}}, \\
    \nonumber \{F_{\alpha,(\beta,\gamma)}^a\}_{\alpha,(\beta,\gamma)} &= \left\{ \langle a | \sigma_\gamma^{R} |a\rangle \, \sigma_\alpha^{A_{\mathrm{i}}} \otimes \sigma_\beta^{A_{\mathrm{o}}} \right\}_{\alpha\geq 0,(\beta,\gamma) \neq (0,0)}.
\end{align}
For the generalized Gell--Mann basis, the symmetric and antisymmetric off-diagonal basis elements of $R$ satisfy
\[
    \langle a | \sigma_\gamma^{R} |a\rangle = 0,
\]
for every $a$. Consequently, these basis elements do not contribute after projection onto the computational basis of the outcome register. Only the identity and the diagonal traceless basis elements survive, reflecting the fact that the outcome register is classical. Hence, in what follows, we restrict to
\[
    \sigma^R_\gamma = \sigma^{\mathrm{diag}}_\gamma,
\]
where $\sigma_0^{\mathrm{diag}}$ denotes the normalized identity on the outcome register and $\{\sigma_k^{\mathrm{diag}}\}_{k \geq 1}$ denote its diagonal traceless generalized Gell--Mann basis elements.

It is further convenient to reorganize the remaining basis elements according to the decomposition:
\begin{multline}
    \nonumber \{(\beta,\gamma)\}_{(\beta,\gamma)\neq(0,0)} =\\ \{(j,0)\}_{1\leq j\leq d_{A_{\mathrm{o}}}^2-1} 
     \cup \{(\beta,k)\}_{0\leq\beta\leq d_{A_{\mathrm{o}}}^2-1,1\leq k\leq n_a-1}.
\end{multline}
This leads to the following affine parametrization of the instrument elements:
\begin{align} \label{eq:instrument_element_basis_grouped_0}
    \nonumber C_{a|x} &= 
    \underbrace{\left(\frac{1}{n_a} \sqrt{\frac{d_{A_{\mathrm{i}}}}{d_{A_{\mathrm{o}}}}} + \sum_{k\geq 1} r^x_{0,0,k} \langle a | \sigma_k^{\mathrm{diag}} |a\rangle\right)}_{\text{baseline:  }\sqrt{{d_{A_{\mathrm{i}}}}/{d_{A_{\mathrm{o}}}}}\, q_{a|x}} \, \sigma_0^{A_{\mathrm{i}}} \otimes \sigma_0^{A_{\mathrm{o}}} \\ 
    \nonumber \qquad &+ 
    \underbrace{\sum_{i,k\geq1} 
     r^x_{i,0,k} 
     \langle a | \sigma_k^{\mathrm{diag}} |a\rangle
    \, \sigma_i^{A_{\mathrm{i}}} \otimes \sigma_0^{A_{\mathrm{o}}}}_{\text{effect-like}} \\
    \qquad &+ 
     \underbrace{\sum_{j\geq1, \alpha,\gamma} 
     r^x_{\alpha,j,\gamma} 
     \langle a | \sigma_\gamma^{\mathrm{diag}} |a\rangle
    \, \sigma_\alpha^{A_{\mathrm{i}}} \otimes \sigma_j^{A_{\mathrm{o}}}}_{\text{transformation-like}},
\end{align}
where the three groups of terms are referred to as the baseline, effect-like, and transformation-like sectors, respectively.

The baseline Choi sector represents the map 
\[
    \rho \;\mapsto \; q_{a|x} \operatorname{tr}[\rho] \frac{\mathds{1}_{A_{\mathrm{o}}}}{d_{A_{\mathrm{o}}}},
\]
that is, the completely depolarizing part of the instrument. The input state is discarded, the output is replaced by the maximally mixed state, and the classical outcome $a$ is generated according to the distribution $q_{a|x}$.

Effect-like terms have the same index pattern as the coefficients that are forbidden by the trace-preserving constraint for a single CPTP map. In a quantum instrument, however, these terms are allowed because they cancel upon summation over the classical outcomes. They encode the POVM element associated with the corresponding instrument outcome. More explicitly, defining
\[
    E_{a|x}
    =
    \left(
        \operatorname{tr}_{A_{\mathrm{o}}} [C_{a|x}]
    \right)^\top,
\]
we have:
\begin{align} \label{eq:povm_element_extraction}
    \nonumber p(a|x) &= \operatorname{tr}\left[\Lambda_{a|x}[\rho]\right] \\
    \nonumber &= \operatorname{tr} \left[\operatorname{tr}_{A_{\mathrm{o}}} [C_{a|x} (\rho^\top \otimes \mathds{1}_{A_{\mathrm{o}}})]\right] \\
    \nonumber &= \operatorname{tr} \left[(\operatorname{tr}_{A_{\mathrm{o}}} [C_{a|x}])^\top \rho\right] \\
    &= \operatorname{tr}[E_{a|x} \rho].
\end{align}
When tracing over the output system $A_{\mathrm{o}}$, only the baseline contribution and the terms proportional to $\sigma_i^{A_{\mathrm{i}}} \otimes \sigma_0^{A_{\mathrm{o}}}$ survive. These determine the dependence of the outcome probabilities on the input state and therefore encode the associated POVM.

Finally, the transformation-like sector contains the remaining degrees of freedom of the instrument beyond those determining its normalization and associated POVM. These terms vanish under the partial trace over the output system and therefore do not affect the associated POVM. Instead, they encode the nontrivial input-output transformation implemented by the operation, conditioned on obtaining outcome $a$.

This parametrization uses
\[
    n_x \left[ d_{A_{\mathrm{i}}}^2(n_a-1) + d_{A_{\mathrm{i}}}^2(d_{A_{\mathrm{o}}}^2-1)n_a \right] = n_xd_{A_{\mathrm{i}}}^2 (n_ad_{A_{\mathrm{o}}}^2-1)
\]
free real parameters. In the standard SDP formulation~\eqref{eq:sdp_instrument}, the optimization variables consist of $n_x n_a$ Choi operators, each belonging to a real vector space of dimension $d_{A_{\mathrm{i}}}^2 d_{A_{\mathrm{o}}}^2$. 
The trace-preserving condition
\[
    \operatorname{tr}_{A_{\mathrm{o}}} \sum_a C_{a|x} = \mathds{1}_{A_{\mathrm{i}}}
\]
imposes $n_xd_{A_{\mathrm{i}}}^2$ independent affine constraints. Consequently, the number of degrees of freedom is
\[  
n_xn_ad_{A_{\mathrm{i}}}^2d_{A_{\mathrm{o}}}^2 - n_xd_{A_{\mathrm{i}}}^2 = n_xd_{A_{\mathrm{i}}}^2 (n_ad_{A_{\mathrm{o}}}^2-1),
\]
which coincides exactly with the number of parameters in our construction. Hence, the parametrization is tight while automatically enforcing the trace-preserving constraint.

Finally, setting $d_{A_{\mathrm{o}}}=1$ reduces the construction to a tight parametrization of POVM elements, for which $E_{a|x}=C_{a|x}^\top$. This special case may be useful in other contexts.

\subsection{Sparse SDP formulation and post-processing} \label{ap:ch_postprocessing}

Combining the previous constructions, the optimization over quantum instruments can be written as the following sparse SDP:
\begin{align}
    \underset{\{r^x_{\alpha, \beta, \gamma}\}}{\mathrm{max}}
    \quad&
    b_0 + \sum_x b^x_i r_i^x  \label{eq:sdp_instrument_sparse}
    \\ \nonumber
    \mathrm{s.t.}
    \quad&
    C _{a|x} = F_0^a + \sum r^x_i F^a_i \succeq0, \\ \nonumber
    & r^x_i \in \mathbb{R} \quad \forall x,i, \\ \nonumber
    & F_0^a = \frac{1}{n_a} \sqrt{\frac{d_{A_{\mathrm{i}}}}{d_{A_{\mathrm{o}}}}} \,\sigma_0^{A_{\mathrm{i}}} \otimes \sigma_0^{A_{\mathrm{o}}}, \\ \nonumber
    & \{F^a_i\}_i = \left\{ \langle a | \sigma_\gamma^{\mathrm{diag}} |a\rangle \, \sigma_\alpha^{A_{\mathrm{i}}} \otimes \sigma_\beta^{A_{\mathrm{o}}} \right\}_{\alpha\geq 0,(\beta,\gamma) \neq (0,0)}, \\ \nonumber
    & b_{0} = \sum_{a,x}
    \operatorname{tr}\!\left[
        B_{a|x} F_0^a 
    \right], \\ \nonumber
    & b_{i}^x = \sum_{a}
    \operatorname{tr}\!\left[
        B_{a|x} F_i^a
    \right].
\end{align}
Here, the index $i$ is understood as a collective label for the triples $(\alpha,\beta,\gamma)$.

After solving the SDP numerically, the resulting instrument is corrected, when necessary, to satisfy the positive-semidefinite constraints up to machine precision. For each fixed input $x$, we mix all instrument elements with the uniform white-noise instrument $\{N_{a|x}\}_a$:
\begin{equation*}
    N_{a|x}
    =
    \frac{1}{n_a}\,
    \mathds{1}_{A_{\mathrm{i}}}\otimes\frac{\mathds{1}_{A_{\mathrm{o}}}}{d_{A_{\mathrm{o}}}},
    \qquad
    \operatorname{tr}_{A_{\mathrm{o}}}\sum_a N_{a|x}
    =
    \mathds{1}_{A_{\mathrm{i}}}.
\end{equation*}
We define:
\begin{equation*}
    \widetilde C_{a|x}
    =
    p_x C_{a|x}
    +
    (1-p_x)N_{a|x},
\end{equation*}
using the same $p_x$ for every outcome $a$, thereby preserving the trace-preserving constraint. Let
\[
    \lambda_a
    =
    \lambda_{\min}
    (C_{a|x}),
    \qquad
    \nu
    =
    \frac{1}{n_ad_{A_{\mathrm{o}}}},
\]
where $\nu$ is the smallest eigenvalue of the white-noise instrument. If $\min_a\lambda_a \geq 0$, we simply choose $p_x=1$. Otherwise, given a target tolerance $\epsilon\geq0$ (in our implementation, $\epsilon \approx 10^{-16}$), we set:
\[
    p_x = \min_{a:\lambda_a<\epsilon}\frac{\nu-\epsilon}{\nu-\lambda_a}
\]
which guarantees that:
\begin{equation*}
    \lambda_{\min}(\widetilde C_{a|x})
    =
    p_x\,\lambda_a+(1-p_x)\,\nu
    \geq
    \epsilon
    \qquad \forall a,
\end{equation*}
with equality attained for at least one outcome. Consequently, $\{\widetilde C_{a|x}\}_a$ is positive semidefinite up to the prescribed tolerance while exactly preserving the trace-preserving constraint.

In this work, the SDPs over quantum instruments were solved using MOSEK~\cite{aps2022mosek, Andersen_2000} through parametrized programs in CVXPY~\cite{diamond2016cvxpy, agrawal2018rewriting}. Parametrized programs avoid recompiling the optimization problem when solving a sequence of closely related SDPs and naturally support warm-starting, which often reduces the number of iterations required for convergence.

\section{Process matrix constraints and parametrization} \label{ap:process_matrix_sdpconstrains}

\subsection{From operational probabilities to the generalized Born rule} \label{ap:from_oper_to_born}

The purpose of this subsection is to make explicit the standard linear-algebraic steps underlying the generalized Born rule \eqref{eq:born_rule} derived in Ref.~\cite{oreshkov_quantum_2012} from operational principles requiring probabilities to behave consistently under classical randomization of local operations and coarse-graining of instrument outcomes. The present subsection does not contain any new results. Rather, we follow the same operational route while making explicit the mathematical extensions that lead from a probability assignment defined only on physical instrument elements to a bilinear functional represented by a Hermitian operator $W$.

Consider two local laboratories, Alice's and Bob's. Upon receiving classical inputs $x$, $y$, they perform local quantum operations and obtain classical outcomes $a$, $b$. For each input, the local operation is described by a quantum instrument, that is, a collection of completely positive (CP) trace-nonincreasing (TNI) maps whose sum is trace preserving (TP). The maps associated to each individual instrument outcome are represented by their Choi operators $C_{a|x}$ and $C_{b|y}$, while the corresponding instruments may be written as
\[
    C_x = \{C_{a|x}\}_a, \qquad C_y = \{C_{b|y}\}_b.
\]
See Appendix~\ref{ap:choi_matrix_sdpconstrains} for a review of the Choi representation and instrument constraints. 

Operationally, the same CPTNI map may appear as part of different instruments. The most general probability assignment could therefore depend not only on the instrument elements $C_{a|x}$ and $C_{b|y}$, but also on the instruments in which they are embedded. In full generality, one may thus write
\begin{equation*}
   p(ab|xy) = P\big((C_{a|x},C_x),(C_{b|y},C_y)\big).
\end{equation*}
Following Ref.~\cite{oreshkov_quantum_2012}, we assume that the probability associated with a local operation depends only on the corresponding CPTNI map and not on the larger instrument context. Equivalently, operationally indistinguishable local operations are assigned the same probability. Under this assumption the probability rule may be written simply as $P(C_{a|x},C_{b|y})$. 

The operational content of the framework enters through two natural consistency requirements. The first is \emph{convex-linearity}. Suppose Alice classically randomizes between two CPTNI maps with Choi operators $C_1$ and $C_2$, selecting $C_1$ with probability $\lambda$ and $C_2$ with probability $1-\lambda$, and subsequently forgets the outcome of the randomization. The resulting event is represented by the Choi operator $\lambda C_1 + (1-\lambda)C_2$. Since the randomization is performed locally and independently of both the underlying process and Bob's operation, the corresponding probability must satisfy:
\begin{equation*}
    P(\lambda C_1 + (1-\lambda)C_2,D) = \lambda P(C_1,D) + (1-\lambda)P(C_2,D),
\end{equation*}
for every CPTNI map $D$ that can occur as an instrument outcome. An analogous relation holds in the second argument.

The second requirement is \emph{additivity under coarse-graining}. If two outcomes belonging to the same instrument are merged into a single outcome, the resulting operation is represented by the sum of their Choi operators. Consequently,
\begin{equation*}
    P(C_1 + C_2,D) = P(C_1,D) + P(C_2,D).
\end{equation*}
A simple consequence is that the null operation has zero probability:
\[
    P(0,D) = 0.
\]
Combining this observation with convex-linearity yields positive homogeneity on the interval $0 \leq t \leq 1$:
\[ 
    P(tC,D) = t\,P(C,D). 
\]

Our aim is to obtain a linear representation of the probability rule. To this end, we first extend $P$ from CPTNI Choi operators to the full cone of CP Choi operators. For the Choi operator $C$ of a CP map, one can choose $\epsilon>0$ sufficiently small such that $\epsilon C$ is CPTNI; for instance, it is enough to take
\[
    \epsilon\leq 1/\lambda_{\max}(\operatorname{tr}_{A_{\mathrm{o}}}[C]),
\]
with $\lambda_{\max}$ denoting the largest eigenvalue and the convention that the bound is arbitrary if $C = 0$. For Choi operators $C$ and $D$ of CP maps, define
\begin{equation*}
    P_+(C,D)\coloneqq \frac{1}{\epsilon\gamma}P(\epsilon C,\gamma D).
\end{equation*}
where $\epsilon C$ and $\gamma D$ are CPTNI. The definition is independent of the particular choices of $\epsilon$ and $\gamma$. Indeed, given two valid choices $\epsilon,\epsilon'$, one may compare both rescalings through $\eta = \min(\epsilon,\epsilon')$ and use homogeneity on the interval $[0,1]$; the same argument applies to the second argument. The map $P_+$ inherits additivity from $P$, and homogeneity extends to arbitrary positive scalars. It is no longer a probability rule on arbitrary CP inputs, and may therefore take values larger than one.

We next extend $P_+$ from the positive cones to the real vector spaces $\mathrm{Herm}(A_{\mathrm{i}}\otimes A_{\mathrm{o}})$ and $\mathrm{Herm}(B_{\mathrm{i}}\otimes B_{\mathrm{o}})$ of Hermitian operators. Any Hermitian $X$ admits a decomposition
\[ 
    X=X_+-X_-, \qquad X_+,X_- \succeq 0, 
\]
for example by its spectral decomposition. For Hermitian $X$ and $Y$, define:
\begin{align*}
    \nonumber \widetilde P(X,Y) &\coloneqq P_+(X_+,Y_+) - P_+(X_+,Y_-)\\
    &- P_+(X_-,Y_+) + P_+(X_-,Y_-).
\end{align*}
Additivity on the positive cone ensures that this definition is independent of the chosen decompositions. Indeed, if
\[ 
    X = X_+-X_- = X'_+-X'_-, 
\] 
then 
\[ 
    X_+ + X'_- = X'_+ + X_-, 
\]
and additivity implies 
\[
    P_+(X_+,D) - P_+(X_-,D) = P_+(X'_+,D) - P_+(X'_-,D)
\]
for every positive $D$. The same argument applies in the second argument. Homogeneity also extends to negative scalars by exchanging the positive and negative parts. Thus $\widetilde P$ is bilinear over $\mathbb{R}$.

The final step is complexification. For
\[ 
    X = X_1 + iX_2, \qquad Y = Y_1 + iY_2, 
\]
with $X_1,X_2,Y_1,Y_2$ Hermitian, define the complex-bilinear extension:
\begin{align*}
    \nonumber \widetilde P_{\mathbb C}(X,Y) &= \widetilde P(X_1,Y_1) - \widetilde P(X_2,Y_2) \\
    &+ i\big(\widetilde P(X_2,Y_1) + \widetilde P(X_1,Y_2)\big).
\end{align*}
This extension has no direct probabilistic interpretation on arbitrary non-Hermitian arguments; it is only the complex-linear extension of the operationally defined rule. On physical Choi operators, which are positive semidefinite and hence Hermitian, it reduces to the original probability assignment.

By the universal property of the tensor product, the complex-bilinear map $\widetilde P_{\mathbb C}$ corresponds to a unique complex-linear functional $\omega$ on 
\[ 
    \mathcal L(A_{\mathrm{i}}\otimes A_{\mathrm{o}}) \otimes \mathcal L(B_{\mathrm{i}}\otimes B_{\mathrm{o}}) 
\] 
such that 
\[ 
    \widetilde P_{\mathbb C}(C_A,C_B) = \omega(C_A\otimes C_B). 
\] 
Since the spaces are finite-dimensional, every such functional can be represented through the Hilbert--Schmidt inner product: there exists an operator $W$ such that 
\[ 
    \omega(X) = \operatorname{tr}[W^\dagger X] 
\] 
for all $X$. Since the probabilities are real on Hermitian physical inputs, $W$ may be chosen Hermitian without changing the probability rule. We therefore obtain the generalized Born rule \eqref{eq:born_rule}. The CPTNI Choi operators generate the Hermitian operator spaces as real vector spaces. After complexification, they span the full operator spaces. Hence the extension is uniquely fixed by the operational probabilities on physical instrument elements.

Non-negativity of probabilities implies
\[
    \operatorname{tr}\!\big[W(C_A\otimes C_B)\big] \geq 0
\]
for all CPTNI Choi operators $C_A$ and $C_B$. By rescaling, this condition extends to all product CP operators. In particular, evaluating it on rank-one product projectors gives 
\[
    \langle\!\langle\psi_{A_{\mathrm{i}}A_{\mathrm{o}}}, \phi_{B_{\mathrm{i}}B_{\mathrm{o}}}|W|\psi_{A_{\mathrm{i}}A_{\mathrm{o}}},\phi_{B_{\mathrm{i}}B_{\mathrm{o}}}\rangle\!\rangle\geq 0
\]
for all vectors $|\psi_{A_{\mathrm{i}}A_{\mathrm{o}}},\phi_{B_{\mathrm{i}}B_{\mathrm{o}}}\rangle\!\rangle \coloneqq \ket{\psi_{A_{\mathrm{i}}A_{\mathrm{o}}}} \otimes \ket{\phi_{B_{\mathrm{i}}B_{\mathrm{o}}}}$. Thus, $W$ is block-positive. If arbitrary shared entangled ancillas between the laboratories are allowed, this requirement strengthens to complete positivity, equivalently:
\begin{equation}\label{eq:posW}
    W\succeq 0,
\end{equation}
as shown in Refs.~\cite{oreshkov_quantum_2012, araujo_witnessing_2015}. The remaining constraints, arising from normalization of probabilities for all local instruments, are discussed in the next subsections.

\subsection{Projector-based SDP formulation of process matrices} \label{ap:proc_projector}

In the previous subsection, we showed that the operational probability rule can be represented by a Hermitian operator $W$ satisfying the generalized Born rule~\eqref{eq:born_rule}. Moreover, non-negativity of probabilities implies \eqref{eq:posW}. We now derive the remaining constraints imposed by probability normalization.

Requiring
\[
    \sum_{a,b} p(ab|xy) = 1
\]
for all local quantum instruments leads to the linear constraints~\cite{araujo_witnessing_2015}:
\begin{align}
    P_{\mathrm{proc}}[W] &= W, \\
    \operatorname{tr}[W] &= d_{A_{\mathrm{o}}}d_{B_{\mathrm{o}}},
    \label{eq:process_matrix_sdpconstrains}
\end{align}
where $P_{\mathrm{proc}}$ is the linear projector
\begin{align} \label{eq:projector_definition}
    \nonumber P_{\mathrm{proc}}[M]
    &\coloneqq{}
    {}_{A_{\mathrm{o}}}M
    +
    {}_{B_{\mathrm{o}}}M
    -
    {}_{A_{\mathrm{o}}B_{\mathrm{o}}}M
    -
    {}_{A_{\mathrm{i}}A_{\mathrm{o}}}M \\
    &\qquad -
    {}_{B_{\mathrm{i}}B_{\mathrm{o}}}M
    +
    {}_{A_{\mathrm{o}}B_{\mathrm{i}}B_{\mathrm{o}}}M
    +
    {}_{A_{\mathrm{i}}A_{\mathrm{o}}B_{\mathrm{o}}}M .
\end{align}
Here we use the trace-and-replace notation defined in Eq.~\eqref{eq:trace_and_replace}. A Hermitian operator $W$ satisfying \eqref{eq:posW}--\eqref{eq:process_matrix_sdpconstrains} is called a \textit{process matrix}.

Equivalently, $P_{\mathrm{proc}}$ may be obtained as the composition of three commuting linear constraints:
\begin{align}
    {}_{A_{\mathrm{i}} A_{\mathrm{o}}}W
    &=
    {}_{A_{\mathrm{i}} A_{\mathrm{o}} B_{\mathrm{o}}}W, \label{eq:process_matrix_constraint_1}
     \\
    {}_{B_{\mathrm{i}} B_{\mathrm{o}}}W
    &=
    {}_{A_{\mathrm{o}} B_{\mathrm{i}} B_{\mathrm{o}}}W, \label{eq:process_matrix_constraint_2}
     \\
    W
    &=
    {}_{A_{\mathrm{o}}}W+{}_{B_{\mathrm{o}}}W-{}_{A_{\mathrm{o}}B_{\mathrm{o}}}W . \label{eq:process_matrix_constraint_3}
\end{align}

Consequently, optimizing a linear functional over bipartite process matrices can be formulated as the SDP:
\begin{align}
    \underset{W}{\mathrm{max}}
    \quad&
    \operatorname{tr}[GW]
    \\ \nonumber
    \mathrm{s.t.}
    \quad&
    W\succeq0,
    \\ \nonumber
    &
    P_{\mathrm{proc}}[W] = W,
    \\ \nonumber
    &
    \operatorname{tr}[W] = d_{A_{\mathrm{o}}}d_{B_{\mathrm{o}}},
    \label{eq:sdp_process_matrix}
\end{align}
where $G$ is the Hermitian operator specifying the objective. In the case of a causal game, $G$ is obtained by summing the payoff-weighted tensor products of the local Choi operators.

\subsection{Sparse basis for the process-matrix subspace}
\label{ap:sparse_process_basis}

\subsubsection{Bipartite scenario}

The original characterization of bipartite process matrices given in Ref.~\cite{oreshkov_quantum_2012} admits a particularly simple affine basis representation. For completeness, we rederive this representation from the projector formulation introduced in Appendix~\ref{ap:proc_projector}.

We employ the same orthonormal Hermitian basis $\{\sigma^X_\alpha\}_{\alpha=0}^{d_X^2-1}$ introduced in Eq.~\eqref{eq:orthonormal_traceless_basis}. Recall that
\begin{equation}
\label{eq:sigma0}
    \sigma^X_0=\frac{\mathds{1}_X}{\sqrt{d_X}},
\end{equation}
while all remaining basis elements are traceless,
\[
    \operatorname{tr}[\sigma^X_i]=0,
    \qquad i\geq1.
\]
As before, Greek indices $\alpha,\beta,\gamma,\delta,\ldots$ run over the full basis, whereas Latin indices $i,j,k,l,\ldots$ are reserved for traceless basis elements.

Expanding the operator $W$ in the corresponding local product basis gives
\begin{equation}
    W
    =
    \sum_{\alpha,\beta,\gamma,\delta}
    w_{\alpha\beta,\gamma\delta}\,
    \sigma^{A_{\mathrm{i}}}_{\alpha}
    \otimes
    \sigma^{A_{\mathrm{o}}}_{\beta}
    \otimes
    \sigma^{B_{\mathrm{i}}}_{\gamma}
    \otimes
    \sigma^{B_{\mathrm{o}}}_{\delta}.
    \label{eq:process_basis_expansion}
\end{equation}

Since trace-and-replace maps retain only the identity component on the subsystems being replaced, the action of the projector constraints is particularly transparent in this basis. For example, the operator ${}_{B_{\mathrm{i}}B_{\mathrm{o}}}W$ retains only basis elements with $(\gamma,\delta)=(0,0)$, whereas ${}_{(1-A_{\mathrm{o}})}W$ retains only basis elements with $\beta>0$, corresponding to traceless components on $A_{\mathrm{o}}$. Because all trace-and-replace maps commute, the process-matrix constraints may be analyzed coefficient-by-coefficient. Using the equivalent constraints \eqref{eq:process_matrix_constraint_1}--\eqref{eq:process_matrix_constraint_3}, one finds that the following coefficients must vanish:
\begin{align}
    {}_{(1-A_{\mathrm{o}})B_{\mathrm{i}}B_{\mathrm{o}}}W=0
    \quad
    &\Longleftrightarrow
    \quad
    w_{\alpha j,00}=0,
    \label{eq:forbidden_process_components_2}
    \\
    {}_{(1-B_{\mathrm{o}})A_{\mathrm{i}}A_{\mathrm{o}}}W=0
    \quad
    &\Longleftrightarrow
    \quad
    w_{00,\gamma l}=0,
    \label{eq:forbidden_process_components_1}
    \\
    {}_{(1-A_{\mathrm{o}})(1-B_{\mathrm{o}})}W=0
    \quad
    &\Longleftrightarrow
    \quad
    w_{\alpha j,\gamma l}=0.
    \label{eq:forbidden_process_components_3}
\end{align}

When all local dimensions are equal to $d$, the total number of forbidden coefficients is
\[
    2d^2(d^2-1)
    +
    d^4(d^2-1)^2
    =
    d^8-2d^6+3d^4-2d^2,
\]
which coincides with the dimension of the null space of $P_{\mathrm{proc}}$. Equivalently, the coefficients that may be nonzero are
\begin{equation}
\label{eq:allowed_process_components}
    w_{\alpha 0,\gamma 0},
    \qquad
    w_{i0,\gamma l},
    \qquad
    w_{\alpha j,k0}.
\end{equation}
All of these correspond to traceless basis elements except for the coefficient $w_{00,00}$, whose value is fixed by the trace normalization condition
\[
    \operatorname{tr}[W]
    =
    d_{A_{\mathrm{o}}}d_{B_{\mathrm{o}}}.
\]
Using $\operatorname{tr}[\sigma_0^X]=\sqrt{d_X}$, one obtains
\[
    w_{00,00}
    =
    \sqrt{
        \frac{
            d_{A_{\mathrm{o}}}d_{B_{\mathrm{o}}}
        }{
            d_{A_{\mathrm{i}}}d_{B_{\mathrm{i}}}
        }
    }.
\]

The total number of allowed coefficients is therefore
\[
    2(d^2-1)^3
    +
    (d^2-1)^2
    +
    2d^2(d^2-1)
    +
    1
    =
    2d^6-3d^4+2d^2,
\]
which equals the rank of $P_{\mathrm{proc}}$. As expected, the dimensions of the allowed and forbidden subspaces sum to the total dimension $d^8$ of the full operator space.

The relative dimension of the process-matrix subspace is therefore
\[
    \frac{\operatorname{rank}P_{\mathrm{proc}}}{d^8}
    =
    \frac{2}{d^2}
    -
    \frac{3}{d^4}
    +
    \frac{2}{d^6},
\]
which vanishes as $d\to\infty$. Thus, for large local dimensions, the space of valid process matrices occupies only a vanishing fraction of the full operator space.

\textit{Interpretation.} The sparse parametrizations of instruments and process matrices admit a simple operational interpretation. To understand the role of different terms, let us examine their contribution to the generalized Born rule:
\begin{align}\label{eq:expandedBorn}
    \nonumber p(ab|xy) &= 
    \sum_{\alpha,\beta,\gamma,\delta} w_{\alpha\beta,\gamma\delta} \operatorname{tr}\!\left[\left(\sigma^{A_{\mathrm{i}}}_{\alpha} \otimes \sigma^{A_{\mathrm{o}}}_{\beta}\right) C_{a|x}\right] \\
    &\qquad \cdot \operatorname{tr}\!\left[\left(\sigma^{B_{\mathrm{i}}}_{\gamma} \otimes \sigma^{B_{\mathrm{o}}}_{\delta}\right) C_{b|y}\right].
\end{align}
For convenience, we reproduce the instrument parametrization of Eq.~\eqref{eq:instrument_element_basis_grouped_0}:
\begin{align} \label{eq:instrument_element_basis_grouped}
    \nonumber C_{a|x} &= 
    \underbrace{\left(\frac{1}{n_a} \sqrt{\frac{d_{A_{\mathrm{i}}}}{d_{A_{\mathrm{o}}}}} + \sum_{k\geq 1} r^x_{0,0,k} \langle a | \sigma_k^{\mathrm{diag}} |a\rangle\right)}_{\text{baseline:  }\sqrt{{d_{A_{\mathrm{i}}}}/{d_{A_{\mathrm{o}}}}}\, q_{a|x}} \, \sigma_0^{A_{\mathrm{i}}} \otimes \sigma_0^{A_{\mathrm{o}}} \\ 
    \nonumber &\qquad + 
    \underbrace{\sum_{i,k\geq1} 
     r^x_{i,0,k} 
     \langle a | \sigma_k^{\mathrm{diag}} |a\rangle
    \, \sigma_i^{A_{\mathrm{i}}} \otimes \sigma_0^{A_{\mathrm{o}}}}_{\text{effect-like}} \\
     &\qquad + 
     \underbrace{\sum_{j\geq1, \alpha,\gamma} 
     r^x_{\alpha,j,\gamma} 
     \langle a | \sigma_\gamma^{\mathrm{diag}} |a\rangle
    \, \sigma_\alpha^{A_{\mathrm{i}}} \otimes \sigma_j^{A_{\mathrm{o}}}}_{\text{transformation-like}}.
\end{align}
An analogous decomposition holds for Bob's instrument elements.

The baseline sector corresponds to a completely depolarizing operation together with a classical probability distribution over outcomes. If both instruments contain only baseline terms, they couple exclusively to the coefficient $w_{00,00}$. In this case the generalized Born rule reduces to
\[
    p(ab|xy) = q_{a|x} q_{b|y},
\]
where $q_{a|x}$ and $q_{b|y}$ are normalized local probability distributions. Thus, the coefficient $w_{00,00}$ describes completely uncorrelated local randomness.

Next consider process matrices containing only coefficients of the form $w_{\alpha 0, \gamma 0}$. These coefficients couple only to the baseline and effect-like sectors of the instruments, since they are orthogonal to all transformation-like components. Let us examine Bob's marginal distribution:
\begin{align*}
    p(b|xy) &= \sum_{\alpha\gamma} w_{\alpha 0, \gamma 0} \underbrace{\operatorname{tr}\left[ (\sigma_\alpha \otimes \sigma_0)(\textstyle\sum_a C_{a|x} ) \right]}_{\sqrt{{d_{A_{\mathrm{i}}}}/{d_{A_{\mathrm{o}}}}} \,\delta_{\alpha0} } \\ 
    &\qquad \cdot \operatorname{tr}\left[ (\sigma_{\gamma} \otimes \sigma_0) C_{b|y} \right].
\end{align*} 
The equality in the underbrace follows from the trace-preserving condition
\[
    \operatorname{tr}_{A_{\mathrm{o}}}
    \left[\sum_a C_{a|x}\right]
    =
    \mathds{1}_{A_{\mathrm{i}}},
\]
together with \eqref{eq:sigma0}. Hence Bob's marginal is independent of Alice's setting $x$. Equivalently, the effect-like sector cancels when summing over Alice's outcomes, while the transformation-like sector is orthogonal to the output-identity process component. By the same argument, Alice's marginal $p(a|xy)$ is independent of Bob's setting $y$. Therefore, coefficients of the form $w_{\alpha0,\gamma0}$ generate only no-signalling, or common-cause, correlations.

Components of the form $w_{\alpha j,k0}$ and $w_{i0,\gamma l}$ couple transformation-like sectors of one party to effect-like sectors of the other. They therefore correspond to components capable of mediating one-way signalling from Alice to Bob and from Bob to Alice, respectively. Let us consider the contribution of the coefficients $w_{i0,\gamma l}$ to \eqref{eq:expandedBorn}:
\begin{equation} \label{eq:signalling_contribution}
    \sum_{i,\gamma,l} w_{i0,\gamma l}
    \underbrace{
    \sum_{k \geq 1}
    r^x_{i,0,k}
    \langle a|\sigma_k^{\mathrm{diag}}|a\rangle
    }_{
    \operatorname{tr}\!\left[
        (\sigma_i^{A_{\mathrm{i}}}\otimes\sigma_0^{A_{\mathrm{o}}})
        C_{a|x}
    \right]
    }
    \underbrace{
    \sum_{\gamma'\geq0}
    r^y_{\gamma,l,\gamma'}
    \langle b|\sigma_{\gamma'}^{\mathrm{diag}}|b\rangle
    }_{
    \operatorname{tr}\!\left[
        (\sigma_\gamma^{B_{\mathrm{i}}} \otimes \sigma_l^{B_{\mathrm{o}}})
        C_{b|y}
    \right]
    }.
\end{equation}
Summing this contribution over Alice's outcome gives zero, because
\[ 
    \sum_a \langle a|\sigma_k^{\mathrm{diag}}|a\rangle = \operatorname{tr}[\sigma_k^{\mathrm{diag}}] = 0 \qquad k \geq 1. 
\]
Thus these terms do not contribute to Bob's marginal distribution and cannot make Bob's marginal depend on Alice's setting $x$. By contrast, summing over Bob's outcome need not give zero, since the transformation-like sector includes terms with $\gamma' = 0$, for which $\sigma_{\gamma'}^{\mathrm{diag}}$ is proportional to the identity. Consequently, Alice's marginal may depend on Bob's instrument, so $w_{i0, \gamma l}$ describes signalling from Bob to Alice. By the same reasoning, components $w_{\alpha j,k0}$ describe signalling from Alice to Bob.

This motivates the interpretation of transformation-like sectors as senders and effect-like sectors as receivers. Such terms do not affect the total normalization, because summing over both outcomes necessarily includes the marginalization over the party carrying the effect-like sector, and the contribution then vanishes. This explains why these components are compatible with the process-matrix constraints.

The forbidden components have a different structure. They couple transformation-like sectors either to baseline sectors or directly to other transformation-like sectors. In such cases, the contribution need not vanish under complete outcome summation and may therefore lead to instrument-dependent normalization factors. For example, a component of the form $w_{00,\gamma l}$ contributes to the total normalization as:
\begin{equation}
    \sum_{ab} p(ab|xy) = \sqrt{\frac{d_{A_{\mathrm{i}}} n_b}{d_{A_{\mathrm{o}}}}}  \sum_{\gamma l} w_{0 0, \gamma l}\, r^y_{\gamma,l,0},
\end{equation}
where $n_b$ is the number of Bob's instrument outcomes. Hence Bob's instrument can affect the total normalization of the probability distribution. Analogous reasoning applies to the other forbidden components.

To summarize, the process-matrix constraints admit transformation-like components only when they are paired with an effect-like component of another laboratory. Such pairings disappear under the appropriate marginalization and therefore preserve normalization. Any coupling of a transformation-like component to only baseline sectors, or directly to another transformation-like component, can contribute to the total outcome sum and hence produce instrument-dependent normalizations. These are precisely the components excluded by the process-matrix constraints. In this sense, transformation-like sectors require an effect-like sector in another laboratory to act as a receiver:
\begin{center}
    \textit{When there is a sender, there must be a receiver.}
\end{center}

\subsubsection{Tripartite scenario}

Although not needed for the numerical investigations of the present work, it is straightforward to extend the sparse-basis construction to the tripartite setting. We therefore briefly record the resulting characterization, as it may be useful in other applications.

Consider three laboratories, Alice, Bob, and Charlie, with incoming and outgoing Hilbert spaces $(A_{\mathrm{i}}, A_{\mathrm{o}})$, $(B_{\mathrm{i}}, B_{\mathrm{o}})$, and $(C_{\mathrm{i}}, C_{\mathrm{o}})$, respectively. Following~\cite{araujo_witnessing_2015}, the projector $P^{\mathrm{tri}}_{\mathrm{proc}}$ onto the tripartite process-matrix subspace is obtained as the composition of the following commuting projectors:
\begin{equation}
\begin{array}{rl}
    P_A[M] & =\prescript{}{\left[1-\left(1-A_{\mathrm{o}}\right) B_{\mathrm{i}} B_{\mathrm{o}} C_{\mathrm{i}} C_{\mathrm{o}}\right] }M, \\[0.3em]
    P_B[M] & =\prescript{}{\left[1-\left(1-B_{\mathrm{o}}\right) A_{\mathrm{i}} A_{\mathrm{o}} C_{\mathrm{i}} C_{\mathrm{o}}\right] }M, \\[0.3em]
    P_C[M] & =\prescript{}{\left[1-\left(1-C_{\mathrm{o}}\right) A_{\mathrm{i}} A_{\mathrm{o}} B_{\mathrm{i}} B_{\mathrm{o}}\right] }M, \\[0.3em]
    P_{A B}[M] & =\prescript{}{\left[1-\left(1-A_{\mathrm{o}}\right)\left(1-B_{\mathrm{o}}\right) C_{\mathrm{i}} C_{\mathrm{o}}\right] }M, \\[0.3em]
    P_{A C}[M] & =\prescript{}{\left[1-\left(1-A_{\mathrm{o}}\right)\left(1-C_{\mathrm{o}}\right) B_{\mathrm{i}} B_{\mathrm{o}}\right] }M, \\[0.3em]
    P_{B C}[M] & =\prescript{}{\left[1-\left(1-B_{\mathrm{o}}\right)\left(1-C_{\mathrm{o}}\right) A_{\mathrm{i}} A_{\mathrm{o}}\right] }M, \\[0.3em]
    P_{A B C}[M] & =\prescript{}{\left[1-\left(1-A_{\mathrm{o}}\right)\left(1-B_{\mathrm{o}}\right)\left(1-C_{\mathrm{o}}\right)\right] }M,
\end{array}
\end{equation} 
We expand $W$ in the local operator basis introduced in \eqref{eq:orthonormal_traceless_basis}:
\begin{equation}
    W
    =
    \sum_{\alpha,\beta,\gamma,\delta,\epsilon,\zeta}
    w_{\alpha\beta,\gamma\delta,\epsilon\zeta}\,
    \sigma^{A_{\mathrm{i}}}_{\alpha}
    \otimes
    \sigma^{A_{\mathrm{o}}}_{\beta}
    \otimes
    \sigma^{B_{\mathrm{i}}}_{\gamma}
    \otimes
    \sigma^{B_{\mathrm{o}}}_{\delta}
    \otimes
    \sigma^{C_{\mathrm{i}}}_{\epsilon}
    \otimes
    \sigma^{C_{\mathrm{o}}}_{\zeta}.
    \label{eq:tripartite_process_basis_expansion}
\end{equation}
As in the bipartite case, the trace-and-replace maps act diagonally in this basis, so the projector constraints can be analyzed coefficient by coefficient.

The single-party constraints $P_A[W] = W$, $P_B[W] = W$, and $P_C[W] = W$ are equivalent to 
\begin{align*}
    \prescript{}{(1-A_{\mathrm{o}})B_{\mathrm{i}}B_{\mathrm{o}}C_{\mathrm{i}}C_{\mathrm{o}}}W &= 0, \\
    \prescript{}{A_{\mathrm{i}}A_{\mathrm{o}}(1-B_{\mathrm{o}})C_{\mathrm{i}}C_{\mathrm{o}}}W &= 0, \\
    \prescript{}{A_{\mathrm{i}}A_{\mathrm{o}}B_{\mathrm{i}}B_{\mathrm{o}}(1-C_{\mathrm{o}})}W &=0,
\end{align*}    
which eliminate the coefficient families
\[
    w_{\alpha j,00,00}, \qquad w_{00,\gamma l,00}, \qquad w_{00,00,\epsilon n}.
\]
Similarly, the two-party constraints $P_{AB}[W] = W$, $P_{AC}[W] = W$, and $P_{BC}[W] = W$ imply
\begin{align*}
    \prescript{}{(1-A_{\mathrm{o}})(1-B_{\mathrm{o}})C_{\mathrm{i}}C_{\mathrm{o}}}W &= 0, \\
    \prescript{}{(1-A_{\mathrm{o}})(1-C_{\mathrm{o}})B_{\mathrm{i}}B_{\mathrm{o}}}W &= 0, \\
    \prescript{}{(1-B_{\mathrm{o}})(1-C_{\mathrm{o}})A_{\mathrm{i}}A_{\mathrm{o}}}W &= 0,
\end{align*}
and therefore remove
\[
    w_{\alpha j,\gamma l,00}, \qquad w_{\alpha j,00,\epsilon n}, \qquad w_{00,\gamma l,\epsilon n}.
\]
Finally, the three-party constraint $P_{ABC}[W] = W$ is equivalent to
\[
    \prescript{}{(1-A_{\mathrm{o}})(1-B_{\mathrm{o}})(1-C_{\mathrm{o}})}W = 0,
\]
which eliminates all coefficients of the form
\[
    w_{\alpha j,\gamma l,\epsilon n}.
\]

Collecting all contributions, the forbidden components are
\[
\begin{array}{ccc}
w_{\alpha j,00,00},
&
w_{00,\gamma l,00},
&
w_{00,00,\epsilon n},
\\[0.6em]
w_{\alpha j,\gamma l,00},
&
w_{\alpha j,00,\epsilon n},
&
w_{00,\gamma l,\epsilon n},
\\[0.6em]
w_{\alpha j,\gamma l,\epsilon n}.
&
&
\end{array}
\]
where $j,l,n \geq 1$ denote traceless output indices. When all local dimensions are equal to $d$, the total number of forbidden coefficients is
\begin{align*}
    3d^2(d^2-1)
    +
    3d^4(d^2-1)^2
    +
    d^6(d^2-1)^3 \\
    =
    d^{12}
    -3d^{10}
    +6d^8
    -7d^6
    +6d^4
    -3d^2.
\end{align*}

The allowed coefficients are precisely those not eliminated by the projector constraints. Grouping them according to their signalling structure yields:
\begin{align*}
    \text{Common cause:} \quad & w_{\alpha 0, \gamma 0, \epsilon 0}, \\ 
    \text{A $\to$ (B or/and C):} \quad & w_{\alpha j, \gamma 0, \epsilon 0} \quad \mbox{{\footnotesize $(\gamma,\epsilon)\neq(0,0)$}}, \\
    \text{B $\to$ (A or/and C):} \quad &  w_{\alpha 0, \gamma l, \epsilon 0} \quad \mbox{{\footnotesize $(\alpha,\epsilon)\neq(0,0)$}},\\
    \text{C $\to$ (A or/and B):} \quad &  w_{\alpha 0, \gamma 0, \epsilon n} \quad \mbox{{\footnotesize $(\alpha,\gamma)\neq(0,0)$}}, \\
    \text{(B and C) $\to$ A:} \quad & w_{i 0, \gamma l, \epsilon n},  \\
    \text{(A and C) $\to$ B:} \quad & w_{\alpha j, k 0, \epsilon n},  \\
    \text{(A and B) $\to$ C:} \quad & w_{\alpha j, \gamma l,m 0}.
\end{align*}
The trace-normalization condition 
\[
    \operatorname{tr}[W] = d_{A_{\mathrm{o}}} d_{B_{\mathrm{o}}} d_{C_{\mathrm{o}}}
\]
fixes the coefficient of the identity basis element:
\[
    w_{00,00,00} = \sqrt{\frac{d_{A_{\mathrm{o}}} d_{B_{\mathrm{o}}} d_{C_{\mathrm{o}}}}{d_{A_{\mathrm{i}}} d_{B_{\mathrm{i}}} d_{C_{\mathrm{i}}}}}. 
\] 
All remaining allowed coefficients correspond to traceless basis elements.

For equal local dimensions $d$, the number of allowed coefficients is
\begin{align*}
    & d^6 + 3(d^6 (d^2-1) - d^2(d^2-1)) + 3 d^4 (d^2-1)^3 \\
    &\quad = 3 d^{10} - 6 d^8 + 7 d^6 - 6 d^4 + 3 d^2.
\end{align*}
Together with the forbidden coefficients counted above, this gives the total number of degrees of freedom $d^{12}$, as required. The relative dimension of the tripartite process-matrix subspace is therefore
\[
    \frac{\operatorname{rank}P^{\text{tri}}_{\mathrm{proc}}}{d^{12}}
    =
    \frac{3}{d^2}-\frac{6}{d^4}+\frac{7}{d^6}-\frac{6}{d^8}+\frac{3}{d^{10}},
\]
which again vanishes in the limit $d\to\infty$. Thus, as in the bipartite case, valid process matrices occupy an asymptotically negligible fraction of the full operator space, with the same leading-order scaling $O(d^{-2})$.

\textit{Interpretation.} The interpretation of the tripartite basis follows the same logic as in the bipartite case. In general, a transformation-like component may contribute consistently only when it is coupled to at least one effect-like component. Operationally, this corresponds to a possible signalling influence from the laboratory carrying the transformation-like sector to the laboratory carrying the effect-like sector.

The tripartite setting introduces a new possibility: two transformation-like sectors may simultaneously couple to a single effect-like sector. Consider, for example, coefficients of the form $w_{i 0, \gamma l, \epsilon n}$. From the perspective of Bob and Charlie alone, the factor
$ (\sigma_\gamma\otimes\sigma_l) \otimes (\sigma_\epsilon\otimes\sigma_n)$ contains transformation-like sectors on both laboratories and would therefore correspond to a forbidden bipartite process component. However, the additional effect-like sector on Alice renders the term consistent. Indeed, upon summing over Alice's outcomes,
\[
    \operatorname{tr} \left[(\sigma_i \otimes\sigma_0)\sum_a C_{a|x}\right] = 0,
\]
since the effect-like contribution vanishes in the outcome sum. Consequently, these coefficients do not contribute to the normalization of the probability distribution.

The same mechanism implies that they disappear whenever Alice is marginalized out. Fixing an instrument for Alice defines the reduced process matrix for Bob and Charlie:
\begin{align*}
    \nonumber &p(bc|xyz) = \operatorname{tr}\left[W^{BC}_{x}(C_{b|y}\otimes C_{c|z})\right] \\
    &\quad \text{where} \quad W^{BC}_{x} = \operatorname{tr}_{A_{\mathrm{i}} A_{\mathrm{o}}}\left[(\textstyle\sum_a C_{a|x}\otimes \mathds{1}_{B_{\mathrm{i}}B_{\mathrm{o}}C_{\mathrm{i}}C_{\mathrm{o}}}) W\right].
\end{align*}
Because the contraction with $\sum_a C_{a|x}$ annihilates the effect-like sector, all coefficients of the form $w_{i0,\gamma l,\epsilon n}$ vanish in $W_x^{BC}$. Thus these terms have no bipartite manifestation: they contribute only to genuine tripartite correlations and disappear under every reduction to an effective bipartite process.

\subsubsection{$N$-partite scenario}

The tripartite construction extends naturally to an arbitrary number of parties. As shown in Ref.~\cite{araujo_witnessing_2015}, for an $N$-partite scenario with local Hilbert spaces $A^1_i, A^1_o, A^2_i, A^2_o, \ldots, A^N_i, A^N_o$ the process-matrix subspace is obtained by imposing one projector constraint for every non-empty subset $\mathcal{X}\subseteq\{1,2,\ldots,N\}$. The corresponding projector is:
\begin{equation}
    P^{\mathcal{X}}_{\mathrm{proc}}[M]={}_{1-\left[\Pi_{k \in \mathcal{X}}\left(1-A_{\mathrm{o}}^k\right) \right] \left[ \Pi_{k \notin \mathcal{X}} A_{\mathrm{i}}^k A_{\mathrm{o}}^k\right]}M,
\end{equation}
and the normalization constraints are equivalently expressed as:
\begin{equation*}
    {}_{\left[\Pi_{k \in \mathcal{X}}\left(1-A_{\mathrm{o}}^k\right) \right] \left[ \Pi_{k \notin \mathcal{X}} A_{\mathrm{i}}^k A_{\mathrm{o}}^k\right]}W=0 \qquad \forall \mathcal{X}\subseteq\{1,2,\ldots,N\}.
\end{equation*}

To characterize the resulting subspace, we expand $W$ in the local operator basis introduced in Eq.~\eqref{eq:orthonormal_traceless_basis}:
\begin{equation*}
    W = \sum_{\alpha_1,\beta_1,\ldots,\alpha_N,\beta_N}w_{\alpha_1\beta_1,\ldots,\alpha_N\beta_N} \bigotimes_{n=1}^{N} \Bigl( \sigma^{A^n_{\mathrm{i}}}_{\alpha_n} \otimes \sigma^{A^n_{\mathrm{o}}}_{\beta_n} \Bigr).
\end{equation*}
As in the bipartite and tripartite cases, each local index pair
$(\alpha_n,\beta_n)$ belongs to one of three sectors:
\begin{align*}
    \mathsf B_n &= \{(0,0)\}, \\
    \mathsf E_n &= \{(i,0): i \geq 1\}, \\
    \mathsf T_n &= \{(\alpha,j): \alpha \geq 0,\ j \geq 1\}.
\end{align*}
These correspond respectively to the baseline, effect-like, and transformation-like sectors.

For a global multi-index $\mu=((\alpha_1,\beta_1),\ldots,(\alpha_N,\beta_N))$, define the sets of parties carrying effect-like and transformation-like components:
\begin{align*}
    E(\mu) &= \{n:(\alpha_n,\beta_n)\in\mathsf{E}_n\}, \\
    T(\mu) &= \{n:(\alpha_n,\beta_n)\in\mathsf{T}_n\}.
\end{align*}
The projector constraints imply a simple general rule: every transformation-like sector must be accompanied by at least one effect-like sector. Equivalently,
\begin{equation*}
    w_\mu = 0 \quad \text{whenever} \quad T(\mu)\neq\varnothing \; \text{and} \; E(\mu) = \varnothing.
\end{equation*}
Thus the forbidden multi-indices are:
\begin{equation*}
    \mathcal{F}_N = \{\mu: T(\mu) \neq \varnothing \; \text{and} \; E(\mu) = \varnothing\},
\end{equation*}
while the allowed multi-indices are simply their complement:
\begin{equation*}
    \mathcal{A}_N = \{\mu: T(\mu) = \varnothing \; \text{or} \; E(\mu) \neq \varnothing\}.
\end{equation*}
This characterization admits a simple interpretation. A process component is allowed whenever every transformation-like sector is accompanied by at least one effect-like sector. The allowed components fall into three classes:
\begin{align*}
    \text{All-baseline:} \quad& T(\mu)=\varnothing\; \text{and }\; E(\mu)=\varnothing,  \\
    \text{Common cause:} \quad& T(\mu)=\varnothing\; \text{and }\; E(\mu)\neq\varnothing, \\
    \text{Sender--receiver:} \quad& T(\mu)\neq\varnothing\; \text{and }\; E(\mu)\neq\varnothing .
\end{align*}
The all-baseline component is fixed by the trace normalization
\[
    \operatorname{tr}[W] = \prod_n d_{A^n_{\mathrm{o}}},
\]
which yields
\[
    w_{00,\ldots,00} = \sqrt{\prod_n \frac{d_{A^n_{\mathrm{o}}}}{d_{A^n_{\mathrm{i}}}}}.
\]
All remaining allowed coefficients correspond to traceless basis elements.

For equal local dimensions $d$, it is straightforward to count the allowed and forbidden components. The cardinalities of the local sectors are:
\begin{align*}
    |\mathsf B_n| &= |\mathsf B| = 1, \\
    |\mathsf E_n| &= |\mathsf E| = d^2-1, \\
    |\mathsf T_n| &= |\mathsf T| = d^2(d^2-1).
\end{align*}
where the subscript $n$ may be dropped because all parties are equivalent. Note that
\[
    |\mathsf B| + |\mathsf E| + |\mathsf T| = d^4,
\] 
which equals the number of local basis elements associated with a single party.

To count the forbidden multi-indices, observe that every party must belong either to the baseline sector or to the transformation-like sector, and at least one party must belong to the latter. Therefore,
\begin{align}
    \nonumber |\mathcal{F}_N| &= \sum_{k=1}^N \binom{N}{k} |\mathsf T|^k \\
    \nonumber &= \sum_{k=0}^N \binom{N}{k} |\mathsf T|^k - 1 \\
    \nonumber &= (1 + |\mathsf T|)^N - 1 \\
    &= (1 + d^2(d^2-1))^N - 1,
\end{align}
where the subtraction of $1$ removes the all-baseline component.

To count the allowed multi-indices, let $k$ denote the number of parties carrying an effect-like sector. For fixed $k \geq 1$, there are $\binom Nk$ ways to choose these parties and each contributes $|\mathsf E|$ possible local indices. The remaining $N-k$ parties may independently belong either to the baseline sector or to the transformation-like sector. Hence:
\begin{align}
    \nonumber |\mathcal A_N| & = 1+\sum_{k=1}^N \binom{N}{k} |\mathsf E|^k  \sum_{m=0}^{N-k} \binom{N-k}{m} |\mathsf T|^m \\ 
    \nonumber &= 1+\sum_{k=1}^N \binom{N}{k} |\mathsf E|^k  (1+|\mathsf T|)^{N-k} \\
    \nonumber &= 1+\sum_{k=0}^N \binom{N}{k} |\mathsf E|^k  (1+|\mathsf T|)^{N-k} - (1+|\mathsf T|)^{N} \\
    \nonumber &= (1 + |\mathsf E| + |\mathsf T|)^N - ((1+|\mathsf T|)^{N} - 1) \\
    &= d^{4N} - (1 + d^2(d^2-1))^N + 1.
\end{align}
where the addition by $1$ denotes the all-baseline component.
As expected, $|\mathcal A_N| + |\mathcal F_N| = d^{4N}$, the total dimension of the ambient operator space. For $N=3$, the formula reproduces
\[
    |\mathcal A_3| = 3d^{10} - 6d^8 + 7d^6 - 6d^4 + 3d^2,
\]
in agreement with the previous explicit tripartite calculation.

The allowed multi-indices correspond precisely to the basis elements spanning the image of the process-matrix projector. Therefore, the number of allowed multi-indices equals the rank of the projector. The relative dimension of the process-matrix subspace is thus:
\begin{equation} \label{eq:relative_dimension_N_partite}
    f_\mathcal A(d,N)
    =
    \frac{|\mathcal A_N|}{d^{4N}}
    =
    1
    -
    \left(
        1-\frac{1}{d^2}+\frac{1}{d^4}
    \right)^N
    +
    \frac{1}{d^{4N}}.
\end{equation}
For fixed $N$, we have $\lim_{d\to\infty} f_\mathcal A(d,N) = 0$. Thus, at large local dimension, valid process matrices occupy only a vanishing fraction of the full operator space. Intuitively, the transformation-like sector grows as $d^4$, whereas the effect-like sector grows only as $d^2$, making it increasingly unlikely that a randomly chosen basis element satisfies the requirement that every transformation-like sector be accompanied by an effect-like sector.

In contrast, for fixed $d$, we have $\lim_{N\to\infty} f_\mathcal A(d,N) = 1$. As the number of parties increases, it becomes overwhelmingly likely that a basis element contains at least one effect-like sector, and therefore satisfies the process-matrix constraints.

These asymptotic regimes have practical implications for numerical optimization. When $f_\mathcal A(d,N) \ll 1$, it is advantageous to parametrize the SDP directly in the allowed basis, since the process-matrix subspace is much smaller than the ambient operator space. Conversely, when $f_\mathcal A(d,N)$ approaches unity, it may be more efficient to work in the full basis and impose the forbidden coefficients as linear constraints.

For qubits,
\[
    f_\mathcal A(2,3)\approx 0.46, \qquad f_\mathcal A(2,4)\approx 0.56,
\]
so the forbidden subspace becomes smaller than the allowed one already at four parties. In this regime, imposing the vanishing of forbidden coefficients leads to fewer linear constraints than explicitly parametrizing all allowed components. The case $d=2$, $N=4$ already corresponds to a process matrix of size $256\times256$, while $d=2$, $N=6$ corresponds to a matrix of size $4096\times4096$, which is within the reach of the numerical methods developed in this work.

For qutrits, the analogous threshold occurs at $N \geq 7$. This corresponds to process matrices of size
\[
    3^{14}\times 3^{14} = 4\,782\,969 \times 4\,782\,969,
\]
which is well beyond the range accessible to current implementations.

\subsection{Sparse SDP formulation for process matrices and post-processing} \label{ap:proc_postprocessing}

Let $\mathcal{I}$ denote the set of allowed multi-indices for the scenario under consideration, excluding the all-identity index. The corresponding operators $\{F_\mu\}_{\mu\in\mathcal I}$ form an orthonormal traceless basis for the traceless part of the process-matrix subspace. By choosing the local operator bases to be the generalized Gell--Mann bases of Eq.~\eqref{eq:gellmann_basis}, these basis elements can be made sparse.

The trace normalization fixes the identity component of the process matrix. Hence every operator in the affine process-matrix subspace can be written as:
\[
    W = \gamma \mathds{1} + \sum_{\mu \in \mathcal{I}} w_\mu F_\mu, \qquad \gamma = \frac{\operatorname{tr}[W]}{d_{\rm tot}},
\]
where $d_{\rm tot}$ is the total dimension of the Hilbert space on which $W$ acts. Therefore, optimizing a linear functional over process matrices can be cast as the sparse SDP:
\begin{align} \label{eq:sparse_process_sdp}
    \underset{w_\mu}{\mathrm{max}}
    \quad&
    b_0 + \sum_{\mu\in\mathcal I}
    b_\mu w_\mu
    \\ \nonumber
    \mathrm{s.t.}
    \quad&
    W = \gamma \mathds{1}
    +
    \sum_{\mu\in\mathcal I}
    w_\mu F_\mu
    \succeq0, \\ \nonumber
    & w_\mu \in \mathbb{R} \qquad
    \forall \mu\in\mathcal I.
\end{align}
In this formulation, the linear process constraints and the trace normalization are enforced by construction.

If the numerical solution of~\eqref{eq:sparse_process_sdp} is not positive semidefinite to the desired precision, we post-process it by mixing with the trivial process matrix
\[
     W_{\mathrm{triv}} = \gamma\mathds{1}.
\]
Specifically, we replace
\[
    W
    \;\mapsto\;
    W'
    =
    (1-\lambda)W
    +
    \lambda W_{\mathrm{triv}},
\]
where $\lambda \in [0,1]$ is chosen as the smallest value such that $W' \succeq 0$ up to the prescribed tolerance. Since $W_{\mathrm{triv}}$ belongs to the same affine process subspace and has the same trace as $W$, this post-processing preserves all linear process constraints and the trace normalization.

The dual of~\eqref{eq:sparse_process_sdp} is
\begin{align} \label{eq:sparse_process_sdp_dual}
    \underset{Z}{\mathrm{min}}
    \quad&
    b_0 + \gamma \operatorname{tr}[Z]
    \\ \nonumber
    \mathrm{s.t.}
    \quad&
    \operatorname{tr}[F_\mu Z] = -b_\mu,\\ \nonumber
    & Z \succeq 0.
\end{align}
Here we used that the basis elements $F_\mu$ are Hermitian.

\section{See-saw algorithm} \label{app:see-saw}

Recall that the correlations generated in a process-matrix scenario are given by the generalized Born rule \eqref{eq:born_rule}, where $W$ is a process matrix and $C_{a|x}$, $C_{b|y}$ are the Choi operators of the local instrument elements. In a causal game, one seeks to optimize a linear functional of the resulting correlations, for instance the payoff
\[
    \sum_{abxy} \pi_{abxy} \operatorname{tr}\left[W\left(C_{a|x}\otimes C_{b|y}\right)\right],
\]
This leads to the non-convex optimization problem:
\begin{align}
    \underset{W, C_{a|x}, C_{b|y}}{\mathrm{max}}
    \quad&
    \sum_{abxy} \pi_{abxy} \operatorname{tr}\left[W\left(C_{a|x}\otimes C_{b|y}\right)\right] \\  \nonumber
    \mathrm{s.t.}
    \quad& W \in \mathcal{W}, \\ \nonumber
    & \{C_{a|x}\}_a \in \mathcal{I}_A, \\  \nonumber
    & \{C_{b|y}\}_b \in \mathcal{I}_B, 
\end{align}
where $\mathcal{W}$ denotes the set of valid process matrices (see Appendix \ref{ap:process_matrix_sdpconstrains} for details), while $\mathcal I_A$ and $\mathcal I_B$ denote the sets of valid quantum instruments for Alice and Bob, respectively. The optimization is non-convex because the objective is multilinear in $W$, $C_{a|x}$, and $C_{b|y}$.

The see-saw method~\cite{Werner_2001, Liang_2007, P_l_2010} is a heuristic alternating-optimization procedure. Starting from an initial feasible triple $(W^0, \{C_{a|x}^0\}_a, \{C_{b|y}^0\}_b)$, one iteratively optimizes over one block of variables while keeping the others fixed.

For instance, the first step may optimize over the process matrix while keeping both local instruments fixed:
\begin{equation*}
    W^1 = \arg\max_{W \in \mathcal{W}} \operatorname{tr}\left[W\left(\textstyle\sum_{abxy} \pi_{abxy} C_{a|x}^0\otimes C_{b|y}^0\right)\right].
\end{equation*}
This is an SDP and therefore gives the optimal process matrix for the fixed instruments.

Next, one optimizes over Alice's instruments while keeping $W = W^1$ and Bob's instruments fixed. The objective can be written as:
\[
    \{C_{a|x}^1\}_{a,x}
    =
    \arg\max_{\{C_{a|x}\}_{a,x}\in\mathcal I_A}
    \sum_{a,x}
    \operatorname{tr}
    \!\left[
        C_{a|x}
        B_{a|x}^{(A)}
    \right],
\]
where
\[
    B_{a|x}^{(A)}
    =
    \sum_{b,y}
    \pi_{abxy}\,
    \operatorname{tr}_{B_{\mathrm{i}}B_{\mathrm{o}}}
    \!\left[
        W^1
        \left(
            \mathds{1}_{A_{\mathrm{i}}A_{\mathrm{o}}}
            \otimes
            C_{b|y}^0
        \right)
    \right].
\]
This is again an SDP, now over Alice's instrument variables. An analogous optimization is then performed over Bob's instruments, keeping $W = W^1$ and $\{C_{a|x}\}_{a,x}=\{C_{a|x}^1\}_{a,x}$ fixed.

The three updates are repeated until convergence, for example until the payoff changes by less than a prescribed tolerance between successive iterations. Each individual step is convex, since it optimizes a linear functional over a convex feasible set. However, the full optimization problem is non-convex, and the see-saw procedure is not guaranteed to find the global optimum. In practice, it may converge to a local optimum or stationary point, so we run the algorithm from multiple random initializations.

\section{Details on the SCS algorithm and implementation} \label{app:scs}

The Splitting Conic Solver (SCS)~\cite{ocpb_16, odonoghue_21} solves optimization problems in the canonical conic form:
\begin{align} \label{eq:scs_canonical_form}
    \min  \quad& c^\top x \\ 
    \nonumber \text{s.t.} \quad& \mathcal{A} x + s = b, \quad (x,s) \in (\mathbb{R}^n,\mathcal{K}),
\end{align}
whose dual problem is:
\begin{align} \label{eq:scs_dual_form}
    \max \quad& -b^\top y \\ 
    \nonumber \text{s.t.} \quad& -\mathcal{A}^\top y + r = c, \quad (r,y) \in (\{0\}^n,\mathcal{K}^*),
\end{align}
where $\mathcal{K}$ is a closed convex cone and $\mathcal{K}^*$ denotes its dual.

For the process-matrix SDP in primal form~\eqref{eq:sparse_process_sdp} and dual form~\eqref{eq:sparse_process_sdp_dual}, the correspondence with the SCS canonical form is:
\begin{equation}
\label{eq:constraint_map_def}
\begin{aligned}
    \text{cones:}\quad
    &
    \begin{aligned}[t]
        \mathcal{K} = \mathcal{K}^* = \mathbb{H}^{d_{\mathrm{tot}}}_+,
    \end{aligned} \\[0.4em]
    \text{variables:}\quad
    &
    \begin{aligned}[t]
        x &= (w_\mu)_{\mu\in\mathcal I} \in \mathbb R^{|\mathcal I|}, \\
        y &= Z \in \mathbb H^{d_{\mathrm{tot}}},
    \end{aligned} \\[0.4em]
    \text{linear map:}\quad
    &
    \begin{aligned}[t]
        \mathcal A x &= -\sum_{\mu\in\mathcal I} x_\mu F_\mu, \\
        (\mathcal A^\top Z)_\mu &= -\operatorname{tr}[F_\mu Z],
    \end{aligned} \\[0.4em]
    \text{data:}\quad
    &
    \begin{aligned}[t]
        b &= \gamma \mathds 1, \\
        c &= -(b_\mu)_{\mu\in\mathcal I}, \\
        -b^\top y &= -\gamma\,\operatorname{tr}[Z].
    \end{aligned}
\end{aligned}
\end{equation}
Here $\mathcal{I}$ is the multi-index set of size $n = |\mathcal{I}|$, $d_{\mathrm{tot}}$ is the total Hilbert-space dimension, and $\mathbb H^{d_{\mathrm{tot}}}_+$ denotes the cone of $d_{\mathrm{tot}}\times d_{\mathrm{tot}}$ complex positive semidefinite matrices. The constant offset $b_0$ in the objective function has been omitted, since it does not affect the optimizer. The symbols $\mathcal{A}$ and $\mathcal{A}^\top$ are shorthand for the linear maps defined above.

Since $\{F_\mu\}_{\mu\in\mathcal I}$ forms an orthonormal basis of the traceless process-matrix subspace, we have:
\begin{align} \label{eq:orthonormality_scs}
    \nonumber\left(\mathcal{A}^\top \mathcal{A} x\right)_\mu & = -\operatorname{tr}\!\left[ F_\mu \left( \sum_\nu -x_\nu F_\nu \right) \right] \\
    \nonumber &= \sum_\nu x_\nu \underbrace{\operatorname{tr}[ F_\mu F_\nu]}_{\delta_{\mu\nu}} \\
    &= x_\mu,
\end{align}
and therefore $\mathcal A^\top\mathcal A = \mathbb I$. Similarly,
\begin{align}
    \nonumber\mathcal{A} \mathcal{A}^\top y & = -\sum_\mu (-\operatorname{tr} [F_\mu\, y]) F_\mu \\
    &= P_{\mathrm{proc}}(y) - \frac{\operatorname{tr}[y]}{d_{\mathrm{tot}}}\, \mathds{1}.
\end{align}
The second identity follows because $\{F_\mu\}_{\mu\in\mathcal I}$ is an orthonormal basis of the traceless process-matrix subspace. Thus, $\mathcal{A}\mathcal{A}^\top$ is precisely the orthogonal projector onto the traceless part of that subspace.

In the next subsection we briefly review the algorithm implemented by SCS. We then describe the modifications introduced in our GPU implementation.

\subsection{Overview of the standard SCS algorithm} 

When strong duality holds, optimal primal and dual solutions of Eqs.~\eqref{eq:scs_canonical_form} and~\eqref{eq:scs_dual_form}, denoted by $(x^\star, s^\star, r^\star, y^\star)$, satisfy the Karush--Kuhn--Tucker (KKT) conditions:
\begin{equation} \label{eq:KKT}
    \begin{cases}
        \mathcal{A} x^{\star} + s^{\star} = b, \\[0.2em]
        \mathcal{A}^{\top} y^{\star} + c = r^{\star}, \\[0.2em]
        \left(y^{\star}\right)^{\top} s^{\star} = 0, \\[0.2em]
        x^{\star}\in\mathbb{R}^n, s^{\star} \in \mathcal{K},  r^{\star}\in\{0\}^n, y^{\star} \in \mathcal{K}^*.
    \end{cases}
\end{equation}
The complementarity condition is equivalent to a vanishing primal--dual gap. Indeed,
\begin{align}
    \nonumber \left(y^{\star}\right)^{\top} s^{\star} &= \left(y^{\star}\right)^{\top} (b-\mathcal{A} x^{\star}) \\
    \nonumber &= \left(y^{\star}\right)^{\top}b-\underbrace{\left(\mathcal{A}^\top y^{\star}\right)^{\top}}_{-c^\top} x^{\star} \\
    \nonumber &= c^\top x^{\star} - (-b^\top y^{\star})\\
    &= 0.
\end{align} 
where we used $r^\star=0$.

Directly solving the KKT system can be problematic when the primal or dual problem is infeasible. SCS therefore introduces a homogeneous self-dual embedding. Two nonnegative scalars, $\tau$ and $\kappa$, are added, where $\tau$ homogenizes the affine constraints and $\kappa$ measures violation of optimality. Dropping the superscript ${}^\star$ for readability, the embedded KKT system becomes
\begin{equation}
\begin{cases}
        \mathcal{A} x+s=b \tau, \\[0.2em] 
        \mathcal{A}^{\top} y + c\tau = r,  \\[0.2em] 
        c^\top x + b^\top y = \kappa, \\[0.2em] 
        x\in\mathbb{R}^n, \; s \in \mathcal{K}, \; r\in\{0\}^n, \; y \in \mathcal{K}^*,  \\[0.2em] 
        (\tau,\kappa)\in \mathbb{R}_+^2.
\end{cases}
\end{equation}
This system can be written compactly as the feasibility problem:
\begin{equation}
\label{eq:homogenous_embedding}
\begin{aligned} 
    \text{find} \quad & (u,v) \\
    \text{s.t.} \quad & \underbrace{\begin{bmatrix}0 & \mathcal{A}^{\top} & c \\
    -\mathcal{A} & 0 & b \\
    -c^\top & -b^\top & 0
    \end{bmatrix}}_{Q}\underbrace{\begin{bmatrix}x \\ y \\ \tau
    \end{bmatrix}}_{u}=\underbrace{\begin{bmatrix}r \\ s \\ \kappa
    \end{bmatrix}}_{v}, \\
    & u=(x,y,\tau) \in \mathbb{R}^n \times \mathcal{K}^* \times \mathbb{R}_+, \\
    & v=(r,s,\kappa) \in \{0\}^n \times \mathcal{K} \times \mathbb{R}_+.
\end{aligned}
\end{equation}
The matrix $Q$ should be understood symbolically, since it contains the linear maps $\mathcal{A}: \mathbb{R}^n \rightarrow \mathbb{H}^{d_{\mathrm{tot}}}$ and $\mathcal{A}^\top: \mathbb{H}^{d_{\mathrm{tot}}} \rightarrow \mathbb{R}^n$. For a numerical implementation, these operators must be represented in a real vector space. A standard choice is to encode a Hermitian matrix by its diagonal entries together with the real and imaginary parts of its strictly upper-triangular entries:
\begin{equation}
    \mathsf{vec}_{\mathbb R}(X) = \bigl( X_{ii},\;
        \sqrt{2}\operatorname{Re}X_{ij},\;
        \sqrt{2}\operatorname{Im}X_{ij}
    \bigr)_{i<j},
\end{equation}
with inverse map $\mathsf{mat}_{\mathbb{R}}$. The factor $\sqrt{2}$ ensures preservation of the Hilbert--Schmidt inner product: 
\[
    \mathsf{vec}_{\mathbb{R}}(X)^\top \mathsf{vec}_{\mathbb{R}}(Y) = \langle X,Y\rangle \coloneqq \operatorname{tr}[X^\dagger Y].
\]
Defining 
\[
    A = [-\mathsf{vec}_{\mathbb R}(F_\mu)]_\mu,
\] 
one obtains
\[
    \mathsf{vec}_{\mathbb R}(\mathcal{A}x) = Ax, \qquad \mathcal{A}^\top y = A^\top \mathsf{vec}_{\mathbb R}(y),
\]
so that $\mathcal{A}^\top$ is represented by the ordinary matrix transpose $A^\top$. Finally, the cone constraints $s\in\mathcal{K}$ and $y\in\mathcal{K}^*$ reduce to requiring that the Hermitian matrices reconstructed from the corresponding real vectors are positive semidefinite.

\paragraph{Properties of the homogeneous self-dual embedding.} 
The homogeneous self-dual embedding possesses several useful properties. First, the variables $\tau, \kappa \geq 0$ satisfy the complementarity relation
\[
    \tau\kappa = 0,
\]
so they cannot both be strictly positive. Second, the system is homogeneous: if $(u, v)$ is a solution of Eq.~\eqref{eq:homogenous_embedding}, then so is $(\lambda u, \lambda v)$ for every $\lambda > 0$. Finally, the embedding is self-dual, in the sense that the dual of Eq.~\eqref{eq:homogenous_embedding} has the same form as the primal problem.

The values of $(\tau,\kappa)$ characterize the outcome of the optimization. If $\tau > 0$, then necessarily $\kappa = 0$, and $(x/\tau, y/\tau, s/\tau)$ is an optimal primal--dual solution pair. If $\tau = 0$ and $\kappa > 0$, then the embedding yields a certificate of infeasibility (more precisely, either primal infeasibility or dual infeasibility, depending on the associated certificate vector). The remaining case $\tau = \kappa = 0$ is degenerate and, by itself, does not provide information about feasibility. However, if in this case one finds $c^\top x < 0$ or $b^\top y < 0$, then a certificate of infeasibility is again obtained.

The trivial solution $u = v = 0$ always satisfies the embedding, but it carries no information about the original optimization problem and is avoided by the iterative scheme employed by SCS. We refer the reader to Ref.~\cite{ocpb_16} for a detailed discussion of the homogeneous self-dual embedding and its properties.

\paragraph{Operator splitting.} 
The SCS algorithm solves the homogeneous self-dual embedding \eqref{eq:homogenous_embedding} using an operator-splitting method based on the alternating direction method of multipliers (ADMM), which scales well to large problems. We briefly review the generic ADMM framework before specializing to SCS.

ADMM is designed for optimization problems of the form:
\begin{align} \label{eq:admm_problem_form}
    \min_{x,z} \quad& f(x)+g(z) \\
    \nonumber \text{s.t.} \quad& z=x.
\end{align}
Introducing a dual variable $\lambda \in \mathbb{R}^n$ and a penalty parameter $\rho > 0$, one defines the augmented Lagrangian:
\begin{equation*} 
    \mathcal L_\rho(x,z,\lambda)
    =
    f(x)+g(z)
    +
    \lambda^\top(z-x)
    +
    \frac{\rho}{2}\|z-x\|_2^2,
\end{equation*}
On feasible points, where $x=z$, both the multiplier term and the quadratic penalty vanish. Consequently, minimizing the augmented Lagrangian over the feasible set is equivalent to solving Eq.~\eqref{eq:admm_problem_form}.

A joint minimization over $(x,z)$ is generally no easier than the original problem. ADMM therefore replaces it by alternating minimizations. Starting from an initial point $(x^0,z^0,\lambda^0)$, the updates are:
\begin{align}  \label{eq:admm_updates_x}
    x^{k+1} &= \arg\min_x \mathcal L_\rho(x,z^k,\lambda^k) \nonumber  \\
        &=\arg\min_x
        \left\{
            f(x)
            +
            \frac{\rho}{2}
            \left\|
                x-
                \left(
                    z^k+\frac{\lambda^k}{\rho}
                \right)
            \right\|_2^2
        \right\}, \\ \label{eq:admm_updates_z}
    z^{k+1}
        &= \arg\min_z \mathcal L_\rho(x^{k+1},z,\lambda^k)\nonumber \\
        &=
        \arg\min_z
        \left\{
            g(z)
            +
            \frac{\rho}{2}
            \left\|
                z-
                \left(
                    x^{k+1}-\frac{\lambda^k}{\rho}
                \right)
            \right\|_2^2
        \right\}. \\ \label{eq:admm_updates_l}
        \lambda^{k+1}
        \nonumber &=\lambda^k + \nabla_{\lambda}\mathcal L_\rho(x^{k+1},z^{k+1},\lambda) \\
        &=
        \lambda^k+\rho (z^{k+1}-x^{k+1}).
\end{align}
The expressions in Eqs.~\eqref{eq:admm_updates_x} and
\eqref{eq:admm_updates_z} are obtained by completing the square and discarding terms independent of the optimization variable. Each update may therefore be interpreted as a proximal step: the variable is chosen to balance minimization of the objective with proximity to the current iterate of the other variable.

Eq.~\eqref{eq:admm_updates_l} performs gradient ascent on the dual variable. The update accumulates the primal residual $z^{k+1}-x^{k+1}$, thereby progressively enforcing the constraint $x=z$ as the iterations proceed. In SCS, this ADMM framework is applied to the homogeneous self-dual embedding \eqref{eq:homogenous_embedding}, where the two alternating steps reduce to projections onto a linear subspace and onto a convex cone, respectively.

\paragraph{Operator splitting for the homogeneous embedding.} 
To apply ADMM to the homogeneous self-dual embedding
\eqref{eq:homogenous_embedding}, we rewrite the feasibility problem as a consensus problem:
\begin{align} \label{eq:hsde_admm_consensus}
    \min_{\substack{(\tilde u,\tilde v) \\ (u,v)}} \quad&
    I_{\mathcal{Q}}(\tilde u,\tilde v)+
    I_{\mathcal C\times\mathcal C^*}(u,v) \\
    \nonumber \mathrm{s.t.} \quad& (\tilde u,\tilde v)=(u,v),
\end{align}
where 
\begin{equation}\label{eq:graphQ}
    \mathcal{Q} = \{(u,v): Qu = v\}
\end{equation}
is the linear subspace defined by the embedding constraints, $\mathcal{C} = \mathbb{R}^n\times\mathcal{K}^*\times\mathbb R_+$, $\mathcal{C}^* = \{0\}^n\times\mathcal K\times\mathbb R_+$, and $I_S$ denotes the indicator function:
\begin{equation}
    I_S(x)=
    \begin{cases}
        0 & x\in S, \\
        +\infty & x\notin S.
    \end{cases}
\end{equation}

Using the scaled form of ADMM, in which the factor $\rho$ is absorbed into the dual variables, the updates \eqref{eq:admm_updates_x}--\eqref{eq:admm_updates_l} become:
\begin{align} \label{eq:admm_updates_hsde}
    \left(\tilde u^{k+1},\tilde v^{k+1}\right)
    &=
    \Pi_{\mathcal{Q}}
    \left(
        u^k+\lambda^k,\,
        v^k+\mu^k
    \right),
    \\ \label{eq:admm_updates_hsde1}
    \left(u^{k+1}, v^{k+1}\right)
    & = \left(\Pi_{\mathcal C} \left(\tilde u^{k+1}-\lambda^k\right),\Pi_{\mathcal C^*}\left(\tilde v^{k+1}-\mu^k\right)\right),
    \\ \label{eq:admm_updates_hsde2}
    \left(\lambda^{k+1}, \mu^{k+1} \right) &= \left(\lambda^k + u^{k+1} - \tilde u^{k+1}, \mu^k+v^{k+1} - \tilde v^{k+1} \right),
\end{align}
where $\Pi_S$ denotes the Euclidean projection onto the set $S$.

Since both objective functions in Eq.~\eqref{eq:hsde_admm_consensus} are indicator functions, the proximal steps appearing in ADMM reduce exactly to projections. The first update is therefore a projection onto the linear subspace $\mathcal{Q}$, while the second is a projection onto the cone $\mathcal{C} \times \mathcal{C}^*$. Owing to the product structure of this cone, the latter projection factorizes as:
\[
    \Pi_{\mathcal C \times \mathcal C^*}(u,v) = \left(\Pi_{\mathcal C}(u),\Pi_{\mathcal C^*}(v)\right).
\]

\paragraph{Eliminating the dual variables.} 
A useful simplification is that the scaled dual variables can be eliminated entirely. One can show that
\[
    \mu^k = u^k, \qquad \lambda^k = v^k,
\]
for every iteration $k$. This property holds at $k=0$ by initialization.

Assume inductively that it holds at iteration $k$. Then
Eq.~\eqref{eq:admm_updates_hsde} becomes
\[
    (\tilde u^{k+1},\tilde v^{k+1}) = \Pi_{\mathcal Q}(a,a), \qquad a = u^k+v^k.
\]
Since $Q$ is skew-symmetric, the orthogonal complement of the graph \eqref{eq:graphQ} is
\[
    \mathcal{Q}^\perp = \{(v,u): (u,v)\in\mathcal{Q}\}.
\]
Therefore,
\[
    \Pi_{\mathcal Q^\perp}(a,a) = (\tilde v^{k+1},\tilde u^{k+1}).
\]
Using the orthogonal decomposition
\[
    (a,a) = \Pi_{\mathcal Q}(a,a) + \Pi_{\mathcal Q^\perp}(a,a),
\]
we obtain
\[
    \tilde u^{k+1} + \tilde v^{k+1} = u^k + v^k.
\]
Combining this identity with the induction hypothesis and the Moreau decomposition
\[
    \Pi_{\mathcal C^*}(x) = x + \Pi_{\mathcal C}(-x),
\]
one finds that
\[
    \lambda^{k+1} = v^{k+1}, \qquad \mu^{k+1} = u^{k+1},
\]
thereby completing the induction. We refer the reader to
Ref.~\cite{ocpb_16} for the detailed derivation.

The practical consequence is that the dual variables $\lambda$ and $\mu$ need not be stored explicitly. Moreover, the projection onto the dual cone can be recovered from the projection onto $\mathcal{C}$ through the Moreau decomposition, so only a single cone projection must be computed at each iteration. More concretely, using $v^{k+1}=\lambda^{k+1}$ and $v^{k}=\lambda^{k}$, this implies the update rule $v^{k+1}=v^{k}+u^{k+1}-\tilde{u}^{k+1}$. In addition, since we do not need to keep track of $\mu^{k+1}$, together with the previous update rule, it also implies that we do not need to keep track of $\tilde{v}^{k+1}$. Consequently, every iteration consists of one projection onto the linear subspace $\mathcal{Q}$ to compute $\tilde{u}^{k+1}$ and one projection onto the cone $\mathcal{C}$ to compute $u^{k+1}$, together with the update rule for $v^{k+1}$.

\paragraph{Closed form for the affine projection.}
From the previous paragraph, the affine projection step
\eqref{eq:admm_updates_hsde} may be written as
\[
    (\tilde u^{k+1},\tilde v^{k+1}) = \Pi_{\mathcal Q}(a,a), \qquad a = u^k + v^k.
\]
Computing this projection amounts to solving:
\begin{equation}
    \Pi_{\mathcal{Q}}(a,a)= \arg\min_{\{(u, v)|Qu=v\}} \left\{\frac{1}{2}\| u-a\|^2 + \frac{1}{2}\| v-a\|^2\right\}.
\end{equation}
Introducing a Lagrange multiplier $\nu$, the corresponding Lagrangian is
\[
    \mathcal L(u,v,\nu) = \frac{1}{2}\|u-a\|_2^2 + \frac{1}{2}\|v-a\|_2^2 + \nu^\top(Qu-v).
\]
The first-order optimality conditions are:
\begin{align}
    \nabla_u \mathcal{L} &= u - a + Q^\top \nu = 0,  \\
    \nabla_v \mathcal{L} &= v - a - \nu = 0,  \\
    \nabla_\nu \mathcal{L} &= Qu - v = 0. \label{eq:optConstr3}
\end{align}
Since $Q$ is skew-symmetric, $Q^\top = -Q$. Eliminating $v$ and $\nu$ from the above equations yields
\[
    (\mathds{1} + Q)u = a.
\]
The matrix $\mathds{1}+Q$ is always invertible because every eigenvalue of a real skew-symmetric matrix is purely imaginary and therefore cannot equal $-1$. Hence
\[
    u = (\mathds{1} + Q)^{-1}a.
\]

\begin{table*}[t]
    \centering
    \setlength{\tabcolsep}{5pt}
    \begin{tabular}{rrrrrrrrr}
        \toprule
        & & & &
        \multicolumn{2}{c}{$Ax$ time $(m\mathrm{s})$} &
        \multicolumn{2}{c}{$A^\top y$ time $(m\mathrm{s})$} \\
        \cmidrule(lr){5-6}
        \cmidrule(lr){7-8}
        $d$ & $n$ & $m$ & $\frac{\mathrm{nnz}(A)}{mn^2}$
        & \texttt{einsum} & SciPy
        & \texttt{einsum} & SciPy & $P_{\mathrm{proc}}$ ($m\mathrm{s}$) \\
        \midrule
        2 & 16   & 87     & 3.3$\cdot 10^{-2}$ & 0.04  & 0.006   & 0.05  & 0.002  &  0.012 \\
        3 & 81   & 1232   & 2.5$\cdot 10^{-3}$ & 0.06  & 0.01   & 0.10  & 0.014  & 0.13 \\
        4 & 256  & 7455   & 3.6$\cdot 10^{-4}$ & 0.23  & 0.11   & 0.33  & 0.12 & 0.27 \\
        5 & 625  & 29424  & 8.1$\cdot 10^{-5}$ & 0.65  & 0.66   & 1.48  & 0.84  & 2.56 \\
        6 & 1296 & 89495  & 2.3$\cdot 10^{-5}$ & 2.4  & 3.5   & 3.6  & 5.0  & 9.11 \\
        7 & 2401 & 228192 & 8.2$\cdot 10^{-6}$ & 7.4  & 13.1  & 15.7 & 18.5 & 46.1 \\
        8 & 4096 & 512127 & 3.3$\cdot 10^{-6}$ & 18.5 & 40.5  & 36.8 & 55.6 & 64.9 \\
        \bottomrule
    \end{tabular}
    \caption{Runtime comparison for the matrix--vector products $Ax$ and $A^\top y$. For reference, we also report timings for the process projector $P_{\mathrm{proc}}$ of Eq.~\eqref{eq:projector_definition}, implemented using JAX trace-and-replace operations. Times are reported in milliseconds. Here $d$ denotes the local dimension, $n=d^4$ is the process-matrix dimension, and $m$ is the number of SDP constraints. The quantity $\mathrm{nnz}(A)/(mn^2)$ reports the density of the matrix representation of $A$, where $\mathrm{nnz}(A)$ is the number of nonzero elements in the corresponding matrix. Benchmarks were performed on an Apple M1 Pro CPU with 16\,GB of RAM. Each timing is averaged over 10 evaluations, with \texttt{block\_until\_ready()} used to ensure accurate measurement. For small dimensions the explicit sparse-matrix representation is faster, whereas the tensor-contraction implementation based on \texttt{einsum} becomes advantageous for $d > 5$. We also tested the JAX BCOO sparse format, but found it slower than both approaches reported here. } 
    \label{tab:a-at-times}
\end{table*}

\paragraph{Solving the affine projection.}
The previous paragraph shows that the first component of the affine projection satisfies
\[
    \tilde u^{k+1} = (\mathds{1} + Q)^{-1}(u^k + v^k).
\]

To compute this quantity efficiently, let
\[
    a = u^k + v^k,
\]
and introduce the block notation
\[
    h = (c,b)^\top, \qquad u = (u_{xy},u_\tau)^\top, \qquad a = (a_{xy},a_\tau)^\top,
\]
where
\[
    u_{xy} = (u_x,u_y)^\top, \qquad a_{xy} = (a_x,a_y)^\top.
\]
Then the linear system
\[
    (\mathds{1}+Q)u = a
\]
takes the form:
\begin{equation}
    \begin{bmatrix}
        M & h \\
        -h^\top & 1 \\
    \end{bmatrix}
    \begin{bmatrix}
        u_{xy} \\ u_\tau
    \end{bmatrix}
    =
    \begin{bmatrix}
        a_{xy} \\ a_\tau
    \end{bmatrix}, \quad M=\begin{bmatrix}
        \mathds{1} & {A}^\top \\
        -{A} & \mathds{1}
    \end{bmatrix},
\end{equation}
where $A$ denotes the real matrix representation of the linear map $\mathcal{A}$ introduced in Eq.~\eqref{eq:constraint_map_def}.

Eliminating $u_\tau$ yields
\[
    u_{xy} = (M+hh^\top)^{-1}(a_{xy}-a_\tau h).
\]
The Sherman–Morrison–Woodbury formula~\cite{ocpb_16} gives the following expression for a rank-one update of the inverse of $M$:
\begin{equation}
    (M+hh^\top)^{-1} = M^{-1} - \frac{M^{-1} h h^\top M^{-1}}{1+h^\top M^{-1} h}.
\end{equation}
Thus, the affine projection step may be written as:
\begin{widetext}
\begin{align}
    \begin{bmatrix}
        \tilde u_{x}^{k+1}\\
        \tilde u_{y}^{k+1}
    \end{bmatrix} &= \left(\mathds{1} - \frac{(M^{-1} h) h^\top }{1+h^\top (M^{-1} h)}\right)\left(M^{-1} \begin{bmatrix}
        v_x^k+u_x^k\\
        v_y^k+u_y^k
    \end{bmatrix}-
    (v_\tau^k+u_\tau^k)
     (M^{-1} h) \right), \\
    \tilde  u_\tau^{k+1} & = a_\tau + c^\top \tilde u_x^{k+1} + b^\top \tilde u_y^{k+1}.
\end{align}
\end{widetext}
The only remaining task is to apply $M^{-1}$. This is required for two quantities:
\[
    M^{-1}h,
\]
which can be precomputed once since $h$ is constant, and
\[
    z = M^{-1}a,
\]
which depends on the current iterate. The system $Mz = a$ reads:
\begin{equation} \label{eq:main_system_affine}
    \begin{bmatrix}
        \mathds{1} & {A}^\top \\
        -{A} & \mathds{1}
    \end{bmatrix}\begin{bmatrix}
        z_{x}\\
        z_{y}
    \end{bmatrix}=\begin{bmatrix}
        a_{x}\\
        a_{y}
    \end{bmatrix}.
\end{equation}
This system may be solved either by a direct factorization or by an iterative method. In this work, we focus on the indirect approach, since it avoids forming large dense matrices and scales more favorably to the process-matrix problems considered here.

Eliminating $z_y$ gives
\begin{align} \label{eq:projection_system}
    (\mathds{1}+A^\top A)z_x & = (a_x - A^\top a_y), \\ \label{eq:projection_system2}
    z_y & = a_y + A z_x.
\end{align}
The first equation can be solved with the conjugate-gradient method, which requires only the matrix-vector products $Az$ and $A^\top y$ and never forms $A^\top A$ explicitly. The standard implementation of SCS~3.2.11 provides GPU acceleration for both the direct and indirect methods used to solve Eq.~\eqref{eq:main_system_affine}.

\paragraph{Final iterates.} 
Combining the previous results yields the iteration used by SCS:
\begin{align} \label{eq:final_iterates}
    \tilde u^{k+1} & = (\mathds{1} + Q)^{-1}(u^k+v^k), \\ \label{eq:final_iterates2}
    u^{k+1} & = \Pi_{\mathcal C}(\tilde u^{k+1}-v^k),  \\ \label{eq:final_iterates3}
    v^{k+1} & = v^k - \tilde u^{k+1} + u^{k+1}.
\end{align}
The affine projection step is written compactly as $(\mathds{1}+Q)^{-1}$, although in practice it is evaluated through the linear systems derived in the previous subsection.

\paragraph{Convergence criteria and rate.} 
Following Refs.~\cite{ocpb_16, odonoghue_21}, convergence is assessed on the original primal-dual pair rather than on the homogeneous embedding itself. Whenever $\tau^k > 0$, the current iterate is rescaled according to
\[
    x^k = \frac{u_x^k}{\tau^k}, \qquad s^k=\frac{v_s^k}{\tau^k}, \qquad y^k=\frac{u_y^k}{\tau^k},
\]
where $u_x$, $u_y$, and $v_s$ denote the primal variable, dual variable, and primal slack components of the embedded iterate.

The primal residual, dual residual, and duality gap are then:
\begin{align}
    p^k &= \mathcal{A} x^k + s^k - b, \\
    d^k &= \mathcal {A}^\top y^k + c, \\
    g^k &= c^\top x^k + b^\top y^k.
\end{align}
The algorithm terminates once
\begin{align}
\|p^k\|_{\infty} & \leq \epsilon_{\mathrm{abs}}+\epsilon_{\mathrm{rel}} \max \left(\|A x\|_{\infty},\|s\|_{\infty},\|b\|_{\infty}\right), \\
\left\|d^k\right\|_{\infty} & \leq \epsilon_{\mathrm{abs}}+\epsilon_{\mathrm{rel}} \max \left(\left\|A^{\top} y\right\|_{\infty},\|c\|_{\infty}\right), \\
\left|g^k\right| & \leq \epsilon_{\mathrm{abs}}+\epsilon_{\mathrm{rel}} \max \left(\left|c^{\top} x\right|,\left|b^{\top} y\right|\right),
\end{align}
in which case $(x^k,s^k,y^k)$ is returned as an approximately primal-dual optimal solution. If instead $\tau^k \approx 0$ while $\kappa^k > 0$, the iterate provides a certificate of primal or dual infeasibility. Here, $\lVert\cdot\rVert_\infty$ denotes the infinity norm, which corresponds to the entry-wise maximum. Throughout this work, we set $\epsilon=\epsilon_{\mathrm{abs}}=\epsilon_{\mathrm{rel}}$, and quote the common value $\epsilon$.

As a first-order method, SCS typically reaches moderate accuracy in relatively few iterations, but exhibits the slow asymptotic convergence characteristic of ADMM-based schemes. Consequently, it is particularly well suited to large-scale conic programs where moderate accuracy is sufficient, whereas interior-point methods are generally preferable when very high accuracy is required.

\subsection{Our implementation of SCS}

\begin{table*}[t]
    \centering
    \setlength{\tabcolsep}{4pt}
    \begin{tabular}{rrrrrr}
        \toprule
        $d$ & $n$
        & NumPy c128 ($\mathrm{s}$)
        & JAX GPU c128 ($\mathrm{s}$) & JAX GPU c64 ($\mathrm{s}$) & Speedup (c128) \\
        \midrule
        2 & 16
        & $0.001 $
        & $0.023 $
        & $0.003 $ & $\approx$ 0.004$\times$  \\
        3 & 81
        & $0.009 $
        & $0.013 $
        & $0.003 $ & $\approx$ 0.7$\times$ \\
        4 & 256
        & $0.032 $
        & $0.032 $
        & $0.004 $ & $\approx$ 1.0$\times$ \\
        5 & 625
        & $0.41 $
        & $0.06 $
        & $0.01 $ & $\approx$ 6.8$\times$ \\
        6 & 1296
        & $3.96 $
        & $0.30 $
        & $0.05 $ & $\approx$ 13.1$\times$ \\
        7 & 2401
        & $26.12 $
        & $1.36 $
        & $0.28 $ & $\approx$ 19.2$\times$ \\
        8 & 4096
        & $136.10 $
        & $6.33 $
        & $1.21 $ & $\approx$ 21.5$\times$ \\
        \bottomrule
    \end{tabular}
    \caption{Benchmark timings for dense Hermitian eigendecomposition as a function of the local dimension $d$, with matrix dimension $n=d^4$. The labels c128 and c64 correspond to the \texttt{complex128} and \texttt{complex64} data types, respectively. Times are reported in seconds and averaged over five runs. JAX timings were measured using \texttt{block\_until\_ready()}. Benchmarks were performed on an NVIDIA Tesla T4 GPU and an Intel Xeon CPU at 2\,GHz. The final column reports the speedup factor of the GPU c128 implementation relative to the NumPy CPU implementation. For $d=2$ and $d=3$, GPU execution is slower due to launch and data-transfer overheads; for larger dimensions the GPU provides substantial acceleration.}
    \label{tab:eigh-benchmark}
\end{table*}

For the process-matrix SDP formulation considered here, the basis $\{F_\mu\}_{\mu\in\mathcal I}$ is orthonormal. As shown in Eq.~\eqref{eq:orthonormality_scs}, this implies
\[
    A^\top A = I.
\]
Consequently, the linear system Eq.~\eqref{eq:projection_system} simplifies to
\begin{align*}
    z_x &= \frac{1}{2}\bigl(a_x - A^\top a_y\bigr), \\
    z_y &= a_y + A z_x.
\end{align*}
Therefore, the affine projection step requires only one application of $A$ and one application of $A^\top$, and no iterative linear solver is needed.

In our implementation, neither the basis matrices $F_\mu$ nor the operator $A$ are formed explicitly. Since each $F_\mu$ is a normalized Kronecker product of local generalized Gell--Mann matrices, the actions of $\mathcal{A}$ and $\mathcal{A}^\top$ can be evaluated directly as tensor contractions over the local basis tensors using JAX's \texttt{einsum}. In particular, the global tensor-product basis is never constructed explicitly; only the local dense Gell--Mann bases are stored.

An alternative approach is to construct the sparse matrix
representation of $A$ and perform explicit sparse matrix--vector multiplications. We found this approach to be less efficient in the parameter regime relevant to this work. A comparison of the different implementations is given in Table~\ref{tab:a-at-times}. We also tested JAX's BCOO sparse matrix format, but found it to be considerably slower than both the explicit sparse implementation and the \texttt{einsum} approach.

More sophisticated solvers, such as SCS and cuLoRADS~\cite{han2024acceleratinglowrankfactorizationbasedsemidefinite}, implement specialized CUDA kernels and optimizations for sparse matrix--vector products. In our setting, however, the dominant cost is the projection onto the positive-semidefinite cone, while the affine projection step is comparatively inexpensive. Consequently, the \texttt{einsum}-based implementation provides a favorable balance between simplicity and performance.

The projection $\Pi_{\mathcal{C}}$ corresponds to a projection onto the positive-semidefinite cone. For Hermitian matrices, this is implemented by computing an eigendecomposition and setting all negative eigenvalues to zero. Since this operation is the dominant computational cost of the algorithm, we perform it on a GPU using JAX~\cite{jax2018github}. In Table~\ref{tab:eigh-benchmark}, we compare eigendecomposition timings on CPU and GPU in both double and single precision. For the largest dimensions considered, the GPU provides roughly a twenty-fold speedup, although the exact factor naturally depends on the particular CPU and GPU hardware used.

In addition, our implementation incorporates several features available in SCS 3.2.11, including type-I Anderson acceleration, over-relaxation, warm-starting, and dynamic scaling of the problem data. We refer the reader to Ref.~\cite{ocpb_16} for a detailed description of these techniques.

\begin{figure}
    \centering
\includegraphics[width=\linewidth]{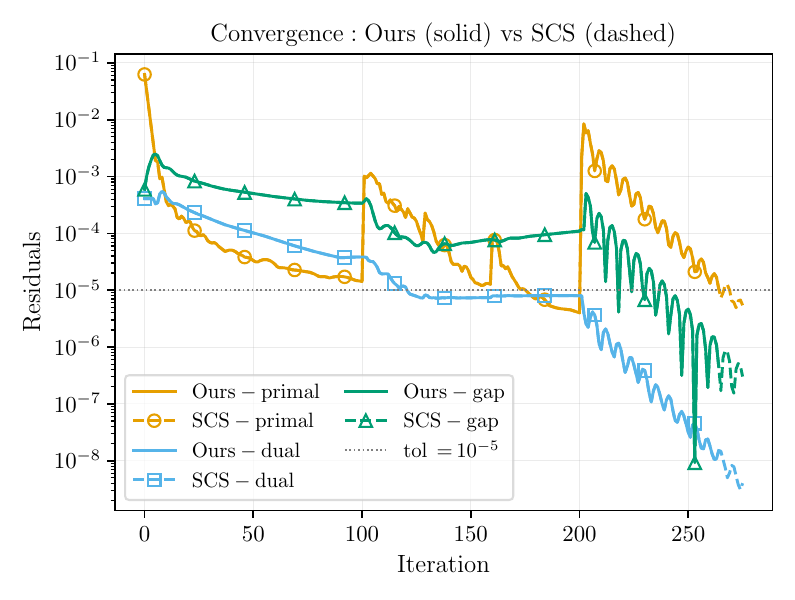}
    \caption{Residual evolution for a representative process-matrix SDP with local dimension $d = 4$. The reference implementation is SCS~3.2.11 with all default settings, except that data normalization is disabled in both solvers. The residual histories agree almost exactly on an iteration-by-iteration basis, confirming that the implementations produce the same ADMM iterates. The discrepancy in the final iteration number is due solely to the termination-check frequency: SCS evaluates convergence every 25 iterations and therefore reports residuals up to iteration 275, whereas our implementation checks convergence every iteration and terminates at iteration 264.}
    \label{fig:residual_matching}
\end{figure}

One notable difference from the standard SCS implementation is that we do not perform full data normalization. Since neither $A$ nor $A^\top$ is formed explicitly, the normalization procedure used in SCS cannot be applied directly. Instead, we employ a partial normalization scheme in which a scale factor $\sigma$ is chosen so that the norms of the cost vector $c$ and the constant constraint vector $b$ coincide. Because this normalization differs from that used in SCS 3.2.11, the resulting residual trajectories are not expected to match exactly. Nevertheless, when normalization is disabled in both implementations, the residuals agree up to numerical precision on an iteration-by-iteration basis, as shown in Fig.~\ref{fig:residual_matching}.

Finally, we support mixed-precision execution. In this mode, a fixed number of initial iterations is performed in single precision before switching to double precision for the remainder of the solve. Empirically, we find that this strategy reaches the same final accuracy while substantially reducing the overall runtime compared with using double precision throughout.

Regarding overall performance, our implementation is approximately twice as slow as native SCS when both are executed on CPU. The tensor-contraction implementations of $A$ and $A^\top$ based on JAX's \texttt{einsum} are individually faster than the corresponding SciPy sparse matrix operations, indicating that these linear maps are not the primary source of the slowdown. Instead, we attribute the performance gap mainly to the overhead of the interpreted Python/JAX solver loop relative to the highly optimized C implementation of SCS.

On GPU hardware, this overhead is more than compensated for once the problem size is sufficiently large. In particular, the projection onto the positive-semidefinite cone (the dominant cost per iteration) can be offloaded to the GPU, allowing the implementation to benefit from the substantial acceleration of dense eigendecompositions. As shown in Table~\ref{tab:wall-clock-time}, the GPU implementation is slower than native SCS for $d\leq 3$, reaches parity around $d = 4$, and becomes increasingly advantageous at larger dimensions, achieving approximately a sixfold speedup at $d = 8$.

This speedup is nevertheless smaller than the roughly twenty-fold acceleration observed for standalone eigendecompositions in Table~\ref{tab:eigh-benchmark}. There are two reasons for this discrepancy. First, the semidefinite-cone projection involves more than the eigendecomposition itself, including matrix assembly, eigenvalue thresholding, matrix reconstruction, and host--device communication. Second, the reference point is native SCS rather than our CPU implementation. Since our CPU code is itself approximately twice as slow as native SCS, part of the GPU acceleration compensates for this software overhead rather than appearing directly as an end-to-end speedup.

\begin{table}[t]
    \centering
    \setlength{\tabcolsep}{5pt}
    \begin{tabular}{rrrr@{\hskip 3pt}rr}
        \toprule
        $d$ & $n$ & iter
        & \multicolumn{2}{c}{SCS (CPU)}
        & {Ours (GPU)} \\
        \cmidrule(lr){4-5}
        & & &  & cone
        &  \\
        \midrule
        2 & 16  & 250 & \SI{37}{\milli\second}  & 76\%  & \SI{3.7}{\second} \\
        3 & 81  & 250 & \SI{577}{\milli\second} & 89\%  & \SI{3.4}{\second} \\
        4 & 256 & 275 & \SI{7.1}{\second}       & 91\%  & \SI{7.1}{\second} \\
        5 & 625 & 275 & \SI{66.4}{\second}        & 93\%  & \SI{21}{\second}  \\
        6 & 1296 & 300 & \SI{8.5}{\minute}      & 95\%  & \SI{2.1}{\minute} \\
        7 & 2401 & 300 & \SI{47.6}{\minute}       & 97\%  & \SI{9.9}{\minute} \\
        8 & 4096 & 300 & \SI{4.4}{\hour}        & 99\%  & \SI{44}{\minute}  \\
        \bottomrule
    \end{tabular}
    \caption{Wall-clock time required to solve a representative process-matrix SDP at each local dimension $d$. The optimization variable is a $d^4\times d^4$ process matrix. A randomly generated feasible pair of instruments defines a GYNI objective, and the problem is solved to tolerance $\epsilon = 10^{-5}$. The column ``iter'' reports the number of ADMM iterations performed by SCS~3.2.11. The column ``cone'' gives the fraction of total CPU runtime spent in the semidefinite-cone projection. For these instances, which lie inside the causal polytope, convergence is typically achieved within a few hundred iterations. Instances attaining genuinely non-causal correlations often require substantially more iterations (typically $10^3$--$3\times10^3$). CPU benchmarks were performed on an Intel Xeon processor at 2\,GHz with 51\,GB of RAM. GPU benchmarks were performed on an NVIDIA Tesla T4 with 15\,GB of memory.}
    \label{tab:wall-clock-time}
\end{table}

\section{Gradient-based joint optimization of causal inequalities} \label{sec:other_methods}

In addition to the SDP-based see-saw method presented in the main text, we also investigated a direct gradient-based optimization approach. The motivation was to optimize the process matrix and local instruments simultaneously, rather than decomposing the problem into the alternating convex subproblems used by the see-saw algorithm. Such methods have proved highly successful in large-scale machine-learning applications and other non-convex optimization settings.

Our implementation was based on the Adam optimizer~\cite{kingma2017adammethodstochasticoptimization} and used JAX~\cite{jax2018github} for automatic differentiation and just-in-time compilation. The main challenge is that process matrices and instruments must satisfy positivity and normalization constraints throughout the optimization. We therefore introduced differentiable parametrizations that enforce these constraints by construction while remaining compatible with gradient-based optimization. Although the resulting methods did not match the performance of the SDP see-saw, we document them here for completeness\footnote{The repository containing the code used for these experiments is available at \url{https://github.com/ecboghiu/ICOJax}.}.

The optimization problem is to maximize the value of a causal inequality over all valid process matrices and local instruments:
\begin{align} \label{eq:nonlinear_opt}
    \underset{W, C_{a|x}, C_{b|y}}{\mathrm{max}} \quad& \sum_{abxy} \pi_{abxy} \operatorname{tr}\left[W\left(C_{a|x}\otimes C_{b|y}\right)\right] \\
    \nonumber \mathrm{s.t.} \quad& W \in \mathcal{W}, \\
    \nonumber & \{C_{a|x}\}_a \in \mathcal{I}_A, \\
    \nonumber & \{C_{b|y}\}_b \in \mathcal{I}_B, 
\end{align}
where $\mathcal W$ denotes the set of valid process matrices (see Appendix~\ref{ap:process_matrix_sdpconstrains}) and $\mathcal I_A$, $\mathcal I_B$ denote the sets of valid instruments for Alice and Bob, respectively (see Appendix~\ref{ap:choi_matrix_sdpconstrains}). To perform joint gradient-based optimization, one requires differentiable parametrizations of these objects that automatically enforce the above constraints. We now describe the parametrizations that we considered.

\paragraph{Parametrization of quantum instruments.}
We investigated two families of differentiable parametrizations for quantum instruments.

The first is based on Stinespring's dilation theorem~\cite{stinespring1955positive}. Any quantum instrument $\Lambda: A_{\mathrm{i}} \to A_{\mathrm{o}} \otimes R$, where $R$ is a classical register storing the measurement outcome, can be represented by an isometry $V^\Lambda: A_{\mathrm{i}} \to A_{\mathrm{o}} \otimes R \otimes E$ into an enlarged Hilbert space containing an environment $E$. The instrument elements are recovered by projecting the register $R$ onto computational-basis states and tracing out the environment, after which the corresponding Choi operators $C_{a|x}$ are constructed. Since every operator obtained in this way arises from a completely positive map, positive semidefiniteness is enforced by construction and no positivity constraints need be imposed during the optimization.

The main drawback of this approach is its poor scaling with dimension. To represent an arbitrary CPTP map, the environment must accommodate the maximal Choi rank, which equals the product of the input and output dimensions of the dilated channel. Since the output space is $A_{\mathrm{o}} \otimes R$, one requires:
\[
    \dim E = \dim A_{\mathrm{i}} \dim A_{\mathrm{o}} \dim R.
\]
As a result, the size of the isometry grows rapidly with the local dimension, making the approach practical only for the smallest instances $d = 2, 3, 4$. Within this framework, we compared several parametrizations of the isometry, including modified Gram--Schmidt orthogonalization, truncated unitaries constructed from Householder reflections, the orthogonal factor of a QR decomposition, and the composite parametrization of the special unitary group introduced in Ref.~\cite{Spengler_2012}. Among these, Householder-reflection and QR-based parametrizations consistently performed best and exhibited similar behavior, whereas the Gram--Schmidt parametrization only rarely located violations of causal inequalities.

The second family is based on the projector characterization of instruments described in Appendix~\ref{ap:choi_matrix_sdpconstrains}. Here we begin with an unconstrained Hermitian matrix whose entries are parametrized by real variables in $[0,1]$. The diagonal entries, together with the real and imaginary parts of the off-diagonal entries, are mapped to the real line via the logit transformation 
\[
    \theta \; \mapsto \; \ln\left[\frac{\theta}{1-\theta}\right].
\]
The trace-preservation constraints are then enforced by a fixed linear projection followed by trace normalization. Individual instrument elements are obtained by projecting the register subsystem $R$ onto computational-basis states $|a \rangle \langle a |$.

This representation is substantially more economical than the dilation-based approach, since the number of optimization variables grows only with the size of the Choi operator itself. However, positivity is no longer guaranteed by construction and must instead be enforced during the optimization, as discussed below\footnote{This parametrization is somewhat redundant. At the time these experiments were carried out, we had not yet derived the sparse instrument parametrization presented in this work. In practice, however, this redundancy is not the dominant limitation of the method. The principal numerical bottleneck is the enforcement of positive semidefiniteness, particularly for the process matrix. We therefore do not expect the tighter instrument parametrization developed in this work to substantially alter the overall performance of the gradient-based approach.}.

\paragraph{Parametrization of process matrices.}
For the process matrix, we explored two complementary parametrizations.

The first mirrors the projector-based instrument parametrization described above. We begin with an unconstrained Hermitian matrix whose entries are generated from logit-transformed parameters, and then apply the process-matrix projector $P_{\mathrm{proc}}$ of Eq.~\eqref{eq:projector_definition}, followed by trace normalization. Recall that $P_{\mathrm{proc}}$ is a fixed linear map consisting of a signed combination of partial traces over the various input and output subsystems, tensored back with appropriately normalized identity operators. This approach enforces all linear process-matrix constraints by construction and is fully differentiable. However, as in the instrument case, the projection does not preserve positive semidefiniteness, so positivity must be enforced separately during the optimization.

The second approach reverses this trade-off. We parametrize the process matrix through a Cholesky-type factorization:
\[
    W = \gamma \frac{LL^\dagger}{\operatorname{tr}[LL^\dagger]}
\]
where $L$ is an unconstrained complex matrix and $\gamma = \operatorname{tr}[W]$ is the fixed process-matrix normalization. This representation guarantees both positive semidefiniteness and the correct trace by construction. The linear process-matrix constraints are no longer automatically satisfied, however, and are instead enforced through a penalty term added to the objective function, which is proportional to $\lVert P_{\mathrm{proc}}[W] - W \rVert$ and measures the distance from the process-matrix subspace.

Neither parametrization simultaneously enforces positivity and the linear process constraints. The projector-based approach guarantees membership in the process-matrix subspace but requires positivity to be corrected during the optimization, whereas the Cholesky parametrization guarantees positivity but only approaches the process-matrix subspace as the penalty weight is increased.

\paragraph{Enforcing positivity and the optimization landscape.}
For the parametrizations that impose the linear constraints by projection (namely, the projector-based instrument parametrization and the corresponding process-matrix parametrization described above), positive semidefiniteness is not guaranteed by construction. Enforcing positivity therefore becomes the central challenge of the gradient-based approach.

We investigated two strategies. The first was an interior-point-inspired barrier method. Specifically, we augmented the objective function with a log-determinant penalty for each Choi operator and for the process matrix. Since $\log\det(X)$ diverges as any eigenvalue of $X$ approaches zero, this penalty should, in principle, keep the iterates inside the interior of the positive semidefinite cone. In practice, however, we found the method to be numerically fragile. The optimization frequently became ill-conditioned, likely because optimal or near-optimal solutions lie close to the boundary of the cone and often have reduced rank. Consequently, the barrier weight required careful tuning: if chosen too small, positivity violations occurred; if chosen too large, the barrier dominated the causal-inequality objective and significantly slowed progress toward high-scoring solutions.

The second strategy, which proved substantially more robust, was to restore positivity explicitly whenever it was violated. Whenever an operator acquired a sufficiently negative eigenvalue, we replaced it by the closest convex mixture with the maximally mixed state that rendered the operator positive semidefinite. Equivalently, we shifted the spectrum just enough to bring the smallest eigenvalue back to zero. This procedure guarantees feasibility while minimally perturbing the current iterate.

Empirically, the projection-based approach produced the best results. For example, in the qubit case $d = 2$, it consistently found causal-inequality violations within approximately $10^{-2}$ of the values obtained by the SDP-based methods. Its main limitation is computational cost: each correction step requires an eigendecomposition, and gradients must be propagated through this eigendecomposition during backpropagation. As a result, the method scales poorly with the local dimension and was ultimately outperformed by the SDP-based see-saw approach.

\paragraph{Direct and neural reparametrizations.}
Using the parametrizations described above, we investigated two variants of the joint gradient-based optimization.

In the first, the unconstrained logit parameters themselves served as the optimization variables and were updated directly using Adam, with positivity enforced through one of the strategies discussed above. In the second, we introduced an additional neural-network layer between the optimizer and the physical parameters. More precisely, a fixed input was mapped through a multilayer perceptron, whose output defined the instrument and process-matrix parameters, while the network weights became the optimization variables. The network was initialized by inversion so that its first forward pass reproduced a valid random starting point. The motivation was that the neural-network parameter space might provide a smoother or better-conditioned optimization landscape than the physical parameters themselves. In practice, however, we did not observe any consistent advantage of the neural parametrization over direct optimization.

More importantly, neither approach was competitive with the SDP-based see-saw method described in the main text. Already for the smallest dimensions $d = 2,3$, the gradient-based methods were both slower and less accurate. The increased computational cost arose primarily from differentiating through repeated eigendecompositions, together with the compilation overhead associated with constructing the process matrix and instrument operators from individual optimization parameters. At the same time, the resulting winning probabilities consistently remained slightly below those obtained by the see-saw algorithm.

The underlying difficulty is precisely the one that the see-saw method avoids by construction. The feasible set is the intersection of a linear subspace with the positive-semidefinite cone. The linear constraints can be enforced efficiently through projection, but positivity remains difficult to handle within a first-order optimization framework. In contrast, the see-saw decomposes the problem into a sequence of convex SDPs, allowing positivity to be enforced exactly at every step. We therefore did not pursue the gradient-based approaches beyond the smallest local dimensions.

\end{document}